# Green and Solid State Reduction of GO Monolayers Sandwiched between Arachidic Acid LB Layers

V. Divakar Botcha[a,*], Pavan K. Narayanam[b,c]

[a]BioSense Institute, University of Novi Sad, Novi Sad, 21000 Serbia

[b]Homi Bhabha National Institute, Training School Complex, Anushaktinagar, Mumbai 400094, India

[c]Materials Chemistry & Metal Fuel Cycle Group, Indira Gandhi Centre for Atomic Research, Kalpakkam 603102, India

*Corresponding author email: divakarbotcha@gmail.com

## Abstract

A novel, single step and environment friendly solid state approach for reduction of graphene oxide (GO)monolayers has been demonstrated, in which, arachidic acid/GO/arachidic acid (AA/GO/AA) sandwich structure obtained by Langmuir-Blodgett (LB) technique was heat treated at moderate temperatures to obtain RGO sheets. Heat treatment of AA/GO/AA sandwich structure at 200 °C results in substantial reduction of GO, with concurrent removal of AA molecules. Such developed RGO sheets possess $sp^2$-C content of ~69%, O/C ratio of ~0.17 and significantly reduced I(D)/I(G) ratio of ~1.1. Ultraviolet photoelectron spectroscopy (UPS) studies on RGO sheets evidenced significant increase in density of states in immediate vicinity of Fermi level and decrease in work function after reduction. Bottom gated field effect transistors fabricated with isolated RGO sheets displayed charge neutrality point at a positive gate voltage, indicating p-type nature, consistent with UPS and electrostatic force microscopy (EFM) measurement results. The RGO sheets obtained by heat treatment of AA/GO/AA sandwich structure exhibited conductivity in the range of 2–7 S/cm and field effect mobility of 0.03–2 $cm^2$/Vs, which are comparable with values reported for RGO sheets obtained by various chemical/thermal reduction procedures. The extent of GO reduction is determined primarily by proximity of AA molecules and found to be unaltered with either escalation of heat treatment temperature or increase of AA content in sandwich structure. The single-step GO reduction approach demonstrated in this work is an effective way for development of RGO monolayers with high structural quality towards graphene-based electronic device applications.

## 1. Introduction

Among diverse graphene-based materials, reduced graphene oxide (RGO) has been extensively explored, in view of it's tunable bandgap, semiconductor characteristics, controllable de-oxygenation and attractive applications as an alternative to graphene [1–3]. RGO layers typically consist of a mixture of $sp^2$-C and $sp^3$-C domains, where the relative ratios of $sp^3$-C and oxygen functional groups are substantially lower than those of GO precursors [4]. RGO, usually obtained by de-oxygenation of GO, was shown to be either semiconducting or conducting, with a range of conductivities depending on the adopted reduction process [1,4–6]. In addition to tunable conductivity, RGO also has several attractive properties, including large specific surface area, high thermal stability, excellent electrical and thermal conductivities, high carrier mobility and superior mechanical properties required for a variety of promising applications in nano-electronics [1,7,8]. Considering these attractive characteristics, utility of RGO monolayers have been widely explored during the past two decades in the fields of optoelectronics, development of photovoltaic devices and sensors [1,7–9].

Several reduction approaches have been explored so far to obtain RGO from a variety of GO structures. These procedures can be mainly classified as 'reduction of GO in solution phase' and 'reduction in the solid state'. In solution phase approaches, as-synthesized GO dispersions were primarily reduced in liquid ambience to form RGO dispersions and the obtained sheets were subsequently transferred onto a variety of solid substrates by suitable deposition methods [10–13]. On the other hand, various solid state reduction approaches were also explored to produce RGO, such as, thermal reduction, exposure to reducing agents/chemical vapors followed by heat treatment [14,15]. These solid state approaches primarily involve heat treatment in the temperature range of 400–1000 °C, either directly or after chemical reduction with reagents such as, hydrazine ($N_2H_4$) derivatives [16], sodium borohydride ($NaBH_4$) [17], alkali (NaOH, KOH) [18], hydrohalic acids (HI) [19], sulfur containing compounds [20] and alcohols [21]. Most of these reducing reagents are toxic in nature and known to be harmful to the environment as well as human beings [22]. Further, usage of these chemicals may lead to generation of by-products and residues along with the RGO, which are not suitable for water, biomedical, agricultural and energy applications [7,9,23,24]. These by-products/residues may also leave corrosive, carcinogenic and toxic constituents on GO sheets during reduction process, which in turn produce non-eco friendly RGO. Exploration of green and environment friendly GO reduction methods not only eliminate the aforementioned issues allied with toxic, corrosive or poisonous by-products, but also facilitate scalable synthesis of hazardous free and sustainable RGO sheets for wide range of opto-electronic device applications. There are other approaches such as microwave, photo-assisted, plasma and electrochemical reductions [19,25–27], which have certain practical concerns and limitations. For example, in the case of electrochemical reduction, the process is largely limited to conductive substrates (electrodes) and cannot facilitate complete elimination of the defects and vacancies inherited from the precursor GO [4,27].

Apart from widely explored chemical and thermal reduction procedures, a few alternative and green reduction methods have also been proposed by various researchers. Pioneering the green reduction approaches, Zhang et al. proposed reduction of GO using ascorbic acid (vitamin C) and demonstrated the efficiency and scalability of the process without any hazardous by-products [28]. Along these lines, various biological reductants including lemon juice, guava leaf, palm oil leaf, green tea extract, barberry fruit etc. were explored for efficient reduction of GO structures [28–32]. Table 1 shows a comparative summary of various green GO reduction methods and the effect of these reducing agents/conditions on degree of reduction, in terms of C/O ratio, $I_D/I_G$ ratio, conductivity and nature/thickness of RGO sheets. It may be noted that most of these green reduction methods were carried out in the solution phase, considering the ease of execution in limited steps. However, adoption of the solution phase processes may induce formation of agglomerates, crumpled or foam structures, multilayer structures and such morphologies are not suitable for the fabrication of electronic devices.

Current manuscript demonstrates a novel, single-step and environment-friendly solid state GO monolayer reduction approach, in which, arachidic acid/GO/arachidic acid (referred as AA/GO/AA, hereafter) sandwich structures developed using Langmuir-Blodgett (LB) technique were heat treated to obtain RGO layers. The LB transferred AA/GO/AA sandwich structure on $SiO_2$/Si substrates consists of a monolayer of AA on either side of GO and, these structures were heat treated at a moderate temperature of 200 °C for AA supported reduction of GO monolayers.

| Reducing agent & method of fabrication | Process of reduction | C/O | I(D)/I(G) | σ (S/m) | Type of layers or thickness of RGO (nm) | References |
|---|---|---|---|---|---|---|
| Ficus carica and Phragmites australis | Solution | 3.57 | - | - | 5-7 layers | [23] |
| Vitamin C | Solution | 5.17 | - | - | multilayers | [33] |
| Brassica oleracea var. gongylodes | Solution | 3.63 | - | - | 40nm | [34] |
| Ziziphus spina-christi | Solution | 1.85-1.92 | | | Multilayers | [35] |
| Ascorbic acid, gallic acid, and sodium citrate | Solution | 4.98, 3.23, 3.53 | 1.93, 1.91, 1.91 | 755.7 | Multilayers | [36] |
| Palm oil leaves extract | Solution | 3 | 1 | - | Multilayers | [37] |
| Vernonia amygdalina plant (locally called "dhebicha") | Solution | 3.88 | - | - | Multilayers | [38] |
| Wood sheets | Solution | - | 1.28 | - | Multilayers | [39] |
| L-ascorbic acid, 50 °C, 48 hr | Solid state | 1.10 | - | - | Multilayers (~ 50 µm) | [40] |

Table 1. A comparative summary of various green reduction methods and their effect on the degree of reduction in terms of C/O ratio, I(D)/I(G) ratio, conductivity and nature/thickness of the RGO sheets.

In order to establish the effectiveness of this novel reduction approach, surface morphology, chemical composition, electronic structures and electrical transport properties of obtained RGO sheets have been investigated. To the best of our knowledge, this is the first green solid state reduction method that produces RGO monolayer sheets. This green and solid state reduction method does not involve usage of any hazardous chemicals and formation of GO aggregates is completely ruled out during this reduction process. The RGO sheets obtained via this procedure are uniformly thick monolayers (~1 nm) with flat morphology, devoid of any wrinkles and suitable for electronic device fabrication. Most importantly, the demonstrated approach doesn't involve/generate any toxic chemicals and the final product (RGO monolayers) are suitable for biomedical, agricultural and energy applications. The current single-step reduction approach is considered as a practical way for development of RGO monolayers with high structural quality, aiming towards graphene-based electronic device applications.

## 2. Experimental details

### 2.1. Development of AA/GO/AA Layered Structures

Langmuir-Blodgett (LB) technique is well known for deposition of organic monolayers and multilayer films of amphiphilic molecules on solid substrates, with an ordered arrangement [41]. Arachidic acid (AA) is an amphiphilic fatty acid molecule that possesses a hydrophilic –COOH group at one end and the hydrophobic hydrocarbon chain at the other end. At the air-water interface, the hydrophilic (polar, –COOH) end is immersed in water, while the hydrophobic chains orient themselves in air, as shown in Fig. S1.

For the LB deposition of arachidic acid molecules, ultra-filtered and de-ionized water (Millipore, 18.2 MΩ-cm) was used as subphase. The LB trough was thoroughly cleaned prior to every deposition where, the trough as well as the barriers was rinsed with acetone as well as ultrapure de-ionized water alternately, and the final rinsing was done with ultrapure de-ionized water. The pH of subphase was maintained at 5.5±0.1 using dilute $HCl/NaHCO_3$. Typically, 150 µl solution of arachidic acid (Sigma Aldrich, 99%) in chloroform (~1 mg/ml) was spread drop by drop on the aqueous subphase with a micro syringe and the subphase temperature was maintained at 10±1 $^{o}$C. The subphase was left idle for 60 min for evaporation of chloroform and stabilization of Langmuir monolayer before compression. A typical surface pressure–mean molecular area (π-A) isotherm of arachidic acid on water surface obtained at a compression speed of 3 mm/min is shown in Fig. S2(a), which clearly shows the phase transitions during the compression of AA molecules. At a typical target pressure of ~30 mN/m, the stable monolayer formed on subphase was transferred onto solid substrates, as shown in the inset of Fig. S2(a). Alternate lifting and dipping of a hydrophilic substrate (at 3 mm/min speed) result in formation of head-to-head and tail-to-tail layered structures, a schematic of which is shown in Figs. S2(b) and S2(c).

In the present work, synthesis of GO was carried out by using modified Hummers-Offeman's method, as described earlier [14]. AA monolayer was transferred initially onto RCA-1 treated $SiO_2$/Si substrates under the above mentioned conditions. Subsequently, GO monolayers were transferred over the AA film to obtain 'AA-GO' structure, under optimized GO transfer conditions described in previous work [42]. Finally, another AA monolayer was transferred over the AA-GO structure to develop a 'AA-GO-AA' layer structure, which will be referred to as AA/GO/AA structure. Along with this, LB deposition of GO layers on a bare $SiO_2$/Si substrate was carried out for the purpose of comparison. In addition, a 3AA/GO/3AA sandwich structure has also been obtained by sequential LB deposition of 3AA layer, GO monolayer and another 3AA layers using similar procedure, for the purpose of comparison studies.

## 2.2. Reduction of GO in AA/GO/AA Layered Structures and characterization

The as-transferred AA/GO/AA layered structure was subjected to heat treatment in vacuum (~$10^{-5}$ mbar) in the temperature in the range of 200–400ºC for 1 hr. For the purpose of comparison, GO monolayers deposited on $SiO_2$/Si (without AA layers) were also subjected to similar heat treatment in vacuum. The morphology, chemical composition, structural properties of as transferred and heat treated sheets were investigated by various techniques and the details are as follows. Surface morphology of GO sheets was studied by Raith-150-Two scanning electron microscope (SEM) at 10 kV and the atomic force microscopy (AFM) images were recorded using Digital Instrument Veeco-Nanoscope IV Multimode scanning probe microscope in tapping mode. The surface electrical characterization of GO and RGO sheets on $SiO_2$/Si substrate was carried out by a Digital Instrument Veeco-Nanoscope IV Multimode scanning probe microscope in dual pass tapping mode. In brief, a Pt-Ir coated tip having a resonance frequency of ~75 kHz and quality factor (k) ~235 was used for these measurements where, topographical information of the sample was obtained in the first scan. The tip was lifted to a height in the range of 15 to 140 nm above the sample surface for the second scan and a DC voltage was applied to the tip for electrostatic force microscopy (EFM) measurements. Chemical composition of the sheets before and after the heat treatment was studied by X-ray photoelectron (XPS) spectroscopy, with Phi500 Versa Probe-II XPS system equipped with monochromatic Al $K_\alpha$ X-ray source. XPS peak fitting was performed using XPS Peak 4.1 software. For Ultraviolet photoelectron spectroscopy (UPS) measurements, a He discharge lamp operated to emit He-I (21.2 eV) and He-II (40.8 eV) flux was used. For UPS data analysis, Avantage V3.9 software was used with Shirley type background fitting and by considering each component peak as Gaussian. The overall energy resolution was 0.7 eV for XPS and 0.1 eV for UPS measurements. For work function measurements by He-I UPS, the sample was biased to -3 V to accelerate low energy secondary electrons. Fourier transform infrared (FT-IR) spectra of these samples on $CaF_2$ substrate were recorded using Perkin Elmer Spectrum One instrument in wavenumber range of 1300–3000 $cm^{-1}$. Micro-Raman spectroscopy was performed using Horiba Jobin Yvon HR800 confocal Raman microprobe equipped with 514 nm $Ar^+$ laser.

Electrical characterization of as-transferred GO and obtained RGO sheets was carried out using a two-probe arrangement, and bottom-gated field effect transistor (FET) geometry was employed to measure the field effect mobility. The device structures were fabricated over isolated RGO monolayer sheets on $SiO_2$(100 nm)/Si substrate, over which, Cr/Au (5 nm/100 nm) source and drain electrodes were patterned by e-beam lithography (Raith-150-Two) and deposited by sputtering. Such Cr/Au electrodes were patterned on RGO monolayers obtained via heat treatment of GO (without AA layers) at 400 $^oC$ or heat treatment of AA/GO/AA sandwich structure at 200 $^oC$, as well. A 150 nm thick aluminium back gate contact was deposited by thermal evaporation. The channel length and width were in the range of 10–20 μm and 10–30 μm, respectively. The device characterization was carried out using a Keithley 4200-SCS semiconductor characterization system. It may be noted that, these electrical characteristics were recorded over several isolated RGO monolayers on a set of FET device structures, prepared under similar conditions and the results obtained from multiple device structures were compiled, compared and only typical results were presented here.

## 3. Results and Discussion

Fig. 1(a) shows a typical SEM image of AA layer deposited on $SiO_2$/Si substrate, which exhibits a uniform coat with rough morphology formed by AA molecules. The SEM image recorded after transfer of GO sheets over the AA layer (Fig. 1(b)) shows uniformly distributed, isolated and clearly visible GO sheets on the rough AA molecule coating. The lateral dimensions of GO sheets are typically in the range of 20–40μm. Further, Fig. 1(c) shows the SEM image of AA layer transferred over the AA/GO structure, in which the GO sheets are not clearly visible due to formation of uniform AA layer overcoat. The AA/GO/AA structure deposited on $SiO_2$/Si substrate was heat treated in vacuum at 200 °C for 1 hr and the subsequently recorded SEM image is shown Fig. 1(d). After heat treatment, the surface modified GO sheets are again clearly visible throughout the substrate, which indirectly suggests the removal of AA molecules during heat treatment and nearly unaffected surface density of GO sheets. It may be noted that, the typical size of these surface modified GO sheets is in the same range of those seen for precursor GO sheets (20-40μm), indicating the nominal effect of heat treatment on overall size and morphology of the GO sheets. Further, the detailed morphological studies carried out via other GO reduction procedures also reported nearly unaltered GO morphology during the reduction process, unless those GO precursors are subjected to harsh chemical or heat treatments [14,15,25,43].

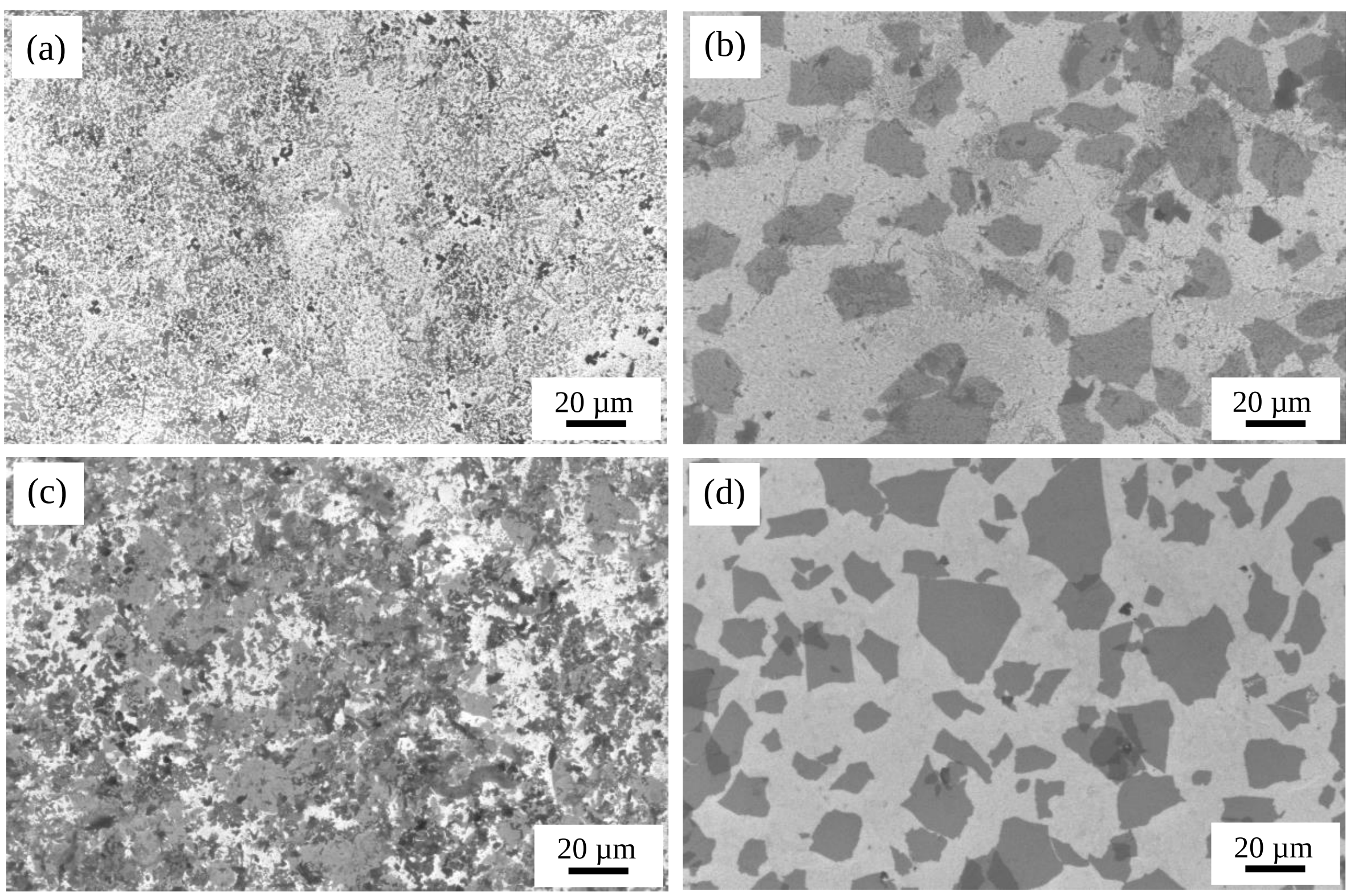

Figure 1. SEM images of as-transferred (a) AA layer, (b) AA/GO and (c) AA/GO/AA layer structure and (d) after heat treatment of AA/GO/AA sandwich structure at 200 °C in vacuum for 1 hr.

The changes in morphological features of AA/GO/AA structure, before and after heat treatment at 200 $^{o}C$ were also examined by recording corresponding AFM images. The typical AFM image of AA/GO/AA layered structure on $SiO_2/Si$ substrate is shown in Fig. 2(a), where the GO sheets appear to be completely encased by AA molecule coat. After heat treatment of the AA/GO/AA structure (Fig. 2(b)), clearly visible GO sheets with good contrast were observed and practically no residual AA was seen at the GO layer surface. However, these sheets exhibited a slightly uneven surface with typical surface roughness ~0.2 nm, which is marginally higher than that of GO/RGO sheets (~0.1 nm) [42]. From both these SEM and AFM studies, it is inferred that the morphology of GO sheets remains nearly unaffected even after heat treatment of AA/GO/AA layered structures at 200 °C, similar to features of GO layers subjected to direct heat treatment at 200 °C (results are not shown here).

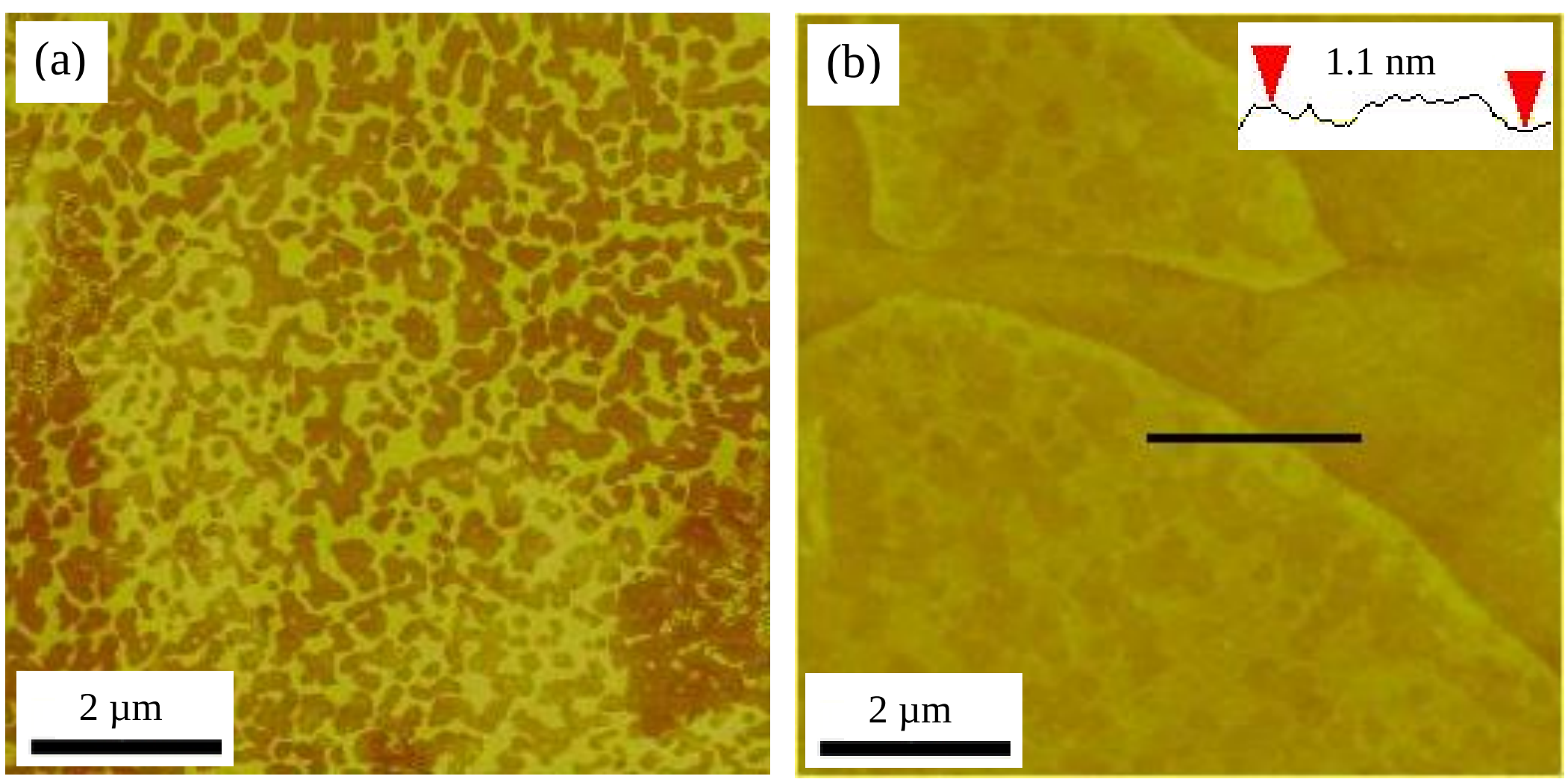

Figure 2. Typical AFM images of as-transferred (a) AA/GO/AA structure on $SiO_2/Si$ and (b) after heat treatment at 200 °C in vacuum for 1hr. Inset of (b) shows the height profile of RGO sheet.

The changes in chemical composition of AA/GO/AA layered structures before and after heat treatment at 200 °C were analyzed by recording the FT-IR spectra in the wavenumber range of 1300–3000 cm$^{-1}$. Fig. 3 shows the FT-IR transmittance spectra of as-transferred AA/GO/AA layered structures on $CaF_2$ substrate and those recorded after subsequent heat treatment. The as-transferred AA/GO/AA layered structure primarily shows dominant AA molecule characteristics where, the intense vibrational bands at ~2918 cm$^{-1}$ and ~2850 cm$^{-1}$ are due to $CH_2$ asymmetric and symmetric stretching modes of hydrocarbon chains of AA molecules, respectively. The intense vibrational band observed at ~1700 cm$^{-1}$ can be attributed to carbonyl stretching vibration of carboxylic acid (–COOH) groups of AA. The weak doublet seen ~1473 cm$^{-1}$ and ~1463 cm$^{-1}$ is due to $CH_2$ scissoring vibrations while, the weak vibrational band at ~2955 cm$^{-1}$ can be ascribed to $CH_3$ asymmetric stretching vibrations of AA molecules [44]. The spectrum recorded after heat treatment at 200 °C shows a substantial decrease in the intensities of both the $CH_2$ vibrational bands (~2918 $cm^{-1}$ and ~2850 $cm^{-1}$) as well as carbonyl stretching band of the carboxylic acid group (~1700 $cm^{-1}$). While the $CH_2$ and –COOH vibrational peaks practically disappear after heat treatment, interestingly a new broad vibrational peak has been found to emerge in the wavenumber range of 1550–1575 $cm^{-1}$. Absence of characteristic vibrational bands of AA molecules after heat treatment at 200 °C indicates nearly complete removal of AA and the fresh peak evolved in the range of 1550–1575 $cm^{-1}$ can be attributed to the skeletal in-plane vibrations of $sp^2$ hybridized C=C bonds in GO sheets [45]. Appearance of prominent vibrational peak allied with $sp^2$ hybridized C=C is a clear indication of the recovery of graphitic carbon content, during heat treatment of GO sheets in close proximity of AA molecules.

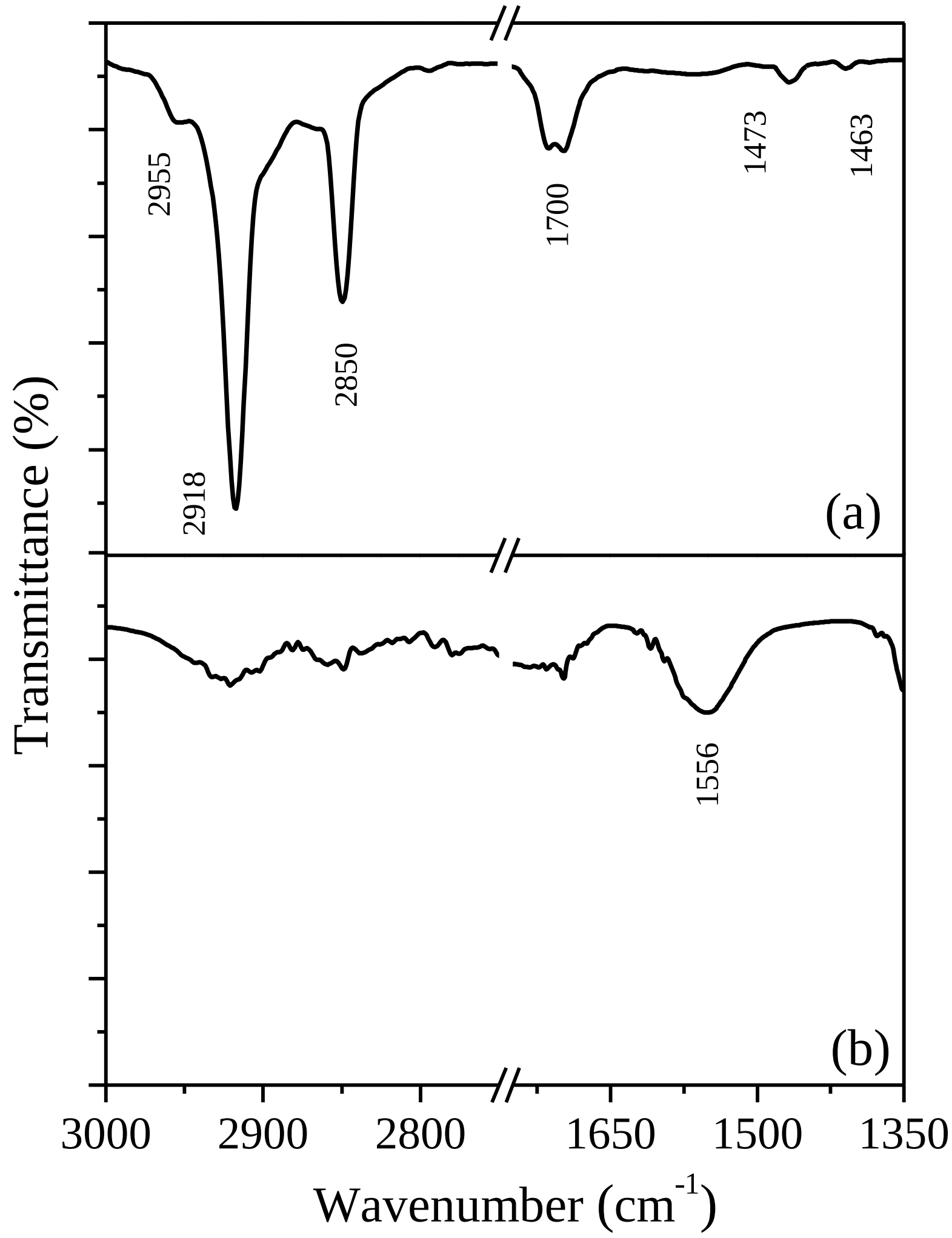

Figure 3. FT-IR spectra of (a) as-transferred AA/GO/AA structures and (b) after heat treatment at 200 °C.

Raman spectra of AA/GO/AA structures transferred on $SiO_2$/Si substrate were recorded before and after heat treatment at 200 °C and the typical spectra are shown in Fig. 4. For the purpose of comparison, Raman spectra of GO sheets on $SiO_2$/Si substrate (without AA layers) and subsequently heat treated at 400 °C were also included in Fig. 4. The Raman spectra of as-transferred GO sheets (Fig. 4(a)) as well as AA/GO/AA layered structure (Fig. 4(c)) typically show the presence of D- and G-bands of GO at ~1340 $cm^{-1}$ and ~1602 $cm^{-1}$, respectively [15,46].

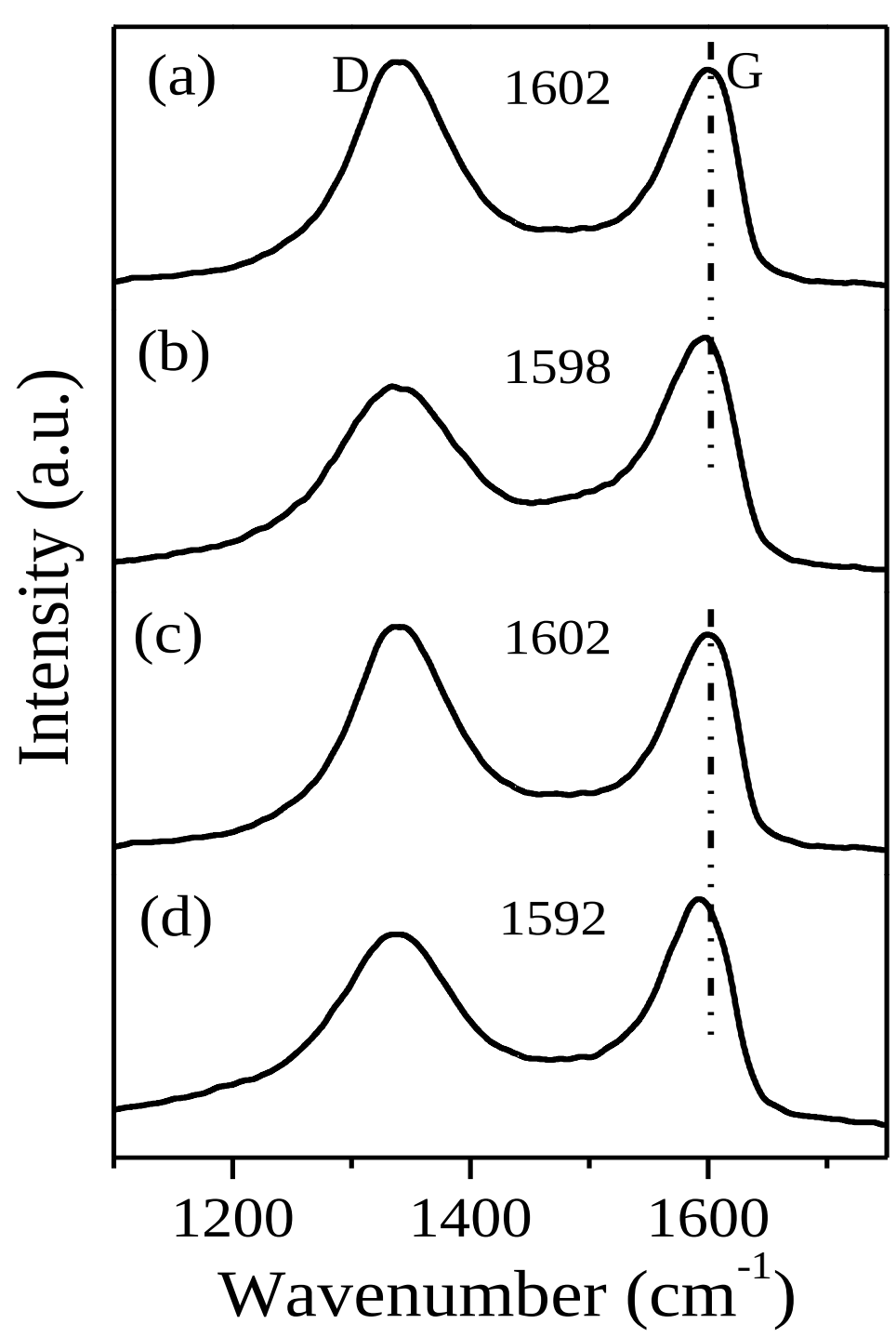

Figure 4. Raman spectra of (a) as-transferred GO/$SiO_2$/Si and (b) after heat treatment at 200 °C, (c) as-transferred AA/GO/AA structure and (d) after heat treatment at 200 °C.

After heat treatment at 200 °C, the GO sheets on $SiO_2$/Si (without AA layers) displayed a small red-shift of G-band to ~1598 $cm^{-1}$ (Fig. 4(b)). In contrast, a much larger red-shift of G-band to ~1592 $cm^{-1}$ has been observed in the case of heat-treated AA/GO/AA structures (Fig. 4(d)). Significant red shift of G-band is an indicative of substantial reduction of GO sheets and restoration of the graphitic network, which is clearly more effective in the case of AA/GO/AA sandwich structure, having AA layers in close proximity of GO sheets. Further, the I(D)/I(G) ratio of as-transferred AA/GO/AA structure was found to be ~1.6, similar to that of GO sheets on $SiO_2$/Si substrate [14,15,47]. It has been observed that the position of G-band and I(D)/I(G) ratio remain nearly same for the GO/AA and AA/GO/AA structures, typically depicting the features similar to those of GO features. However, the I(D)/I(G) ratio of AA/GO/AA sandwich structure was found to reduce to a value of ~1.1, after heat treatment at 200 °C (Fig. 4(d)). Significant decrease in I(D)/I(G) ratio of heat treated AA/GO/AA structures is again an indicative of a substantial reduction of GO and recovery of the graphitic carbon network [8,15]. Thus, the observed major red shift of G-band and appreciable decrease of I(D)/I(G) ratio together evidence substantial reduction and restoration of graphitic network in AA/GO/AA sandwich structure after heat treatment at 200 °C.

The extent of de-oxygenation and the nature of chemical bonding in RGO sheets formed after heat treatment of AA/GO/AA sandwich structure have been studied by recording the corresponding XPS spectra (Fig. 5). For the purpose of comparison, the de-convoluted C1*s* spectra of GO sheets on $SiO_2$/Si substrate (without the AA layer) before and after heat treatment at 400 °C in vacuum are also shown in Fig. 5(a) and 5(b), respectively. The C1*s* spectrum of GO in sandwich structure showed considerably different features from that of as-transferred GO sheets (Fig. 5 (a)) due to the presence of AA layer coat, hence is not shown here. Fig. 5(c) shows the de-convoluted C1*s* core level XPS spectra of RGO sheets obtained by heat treatment of AA/GO/AA structure at 200 °C. The corresponding peak positions and relative integrated intensities of the de-convoluted components are listed in Table 2, along with those of as-transferred and heat treated GO samples. Both the heat treated GO sheets (Fig. 5(b)) and AA/GO/AA structures (Fig. 5(c)) displayed dominant $sp^2$-C content, along with the reduced $sp^3$-C/damaged alternant hydrocarbon and oxygen functional group contributions on the higher binding energy side (> 285.5 eV). From Fig. 5(b) and Table 2, it may be observed that the $sp^2$-C content of RGO sheets obtained by heat treatment at 400 °C (without the AA layers) is ~63 % and the O/C ratio is ~0.24. On the other hand, for the AA/GO/AA structure heat treated at 200 °C, the $sp^2$-C content was found to be ~69% and the O/C ratio is seen at ~0.17. Table 2 also shows that the ratio of non-graphitic to graphitic carbon content 'X' for the heat treated AA/GO/AA structure is significantly lower (~0.41) than those of precursor GO (~1.16) and GO-400 °C (~0.52) samples.

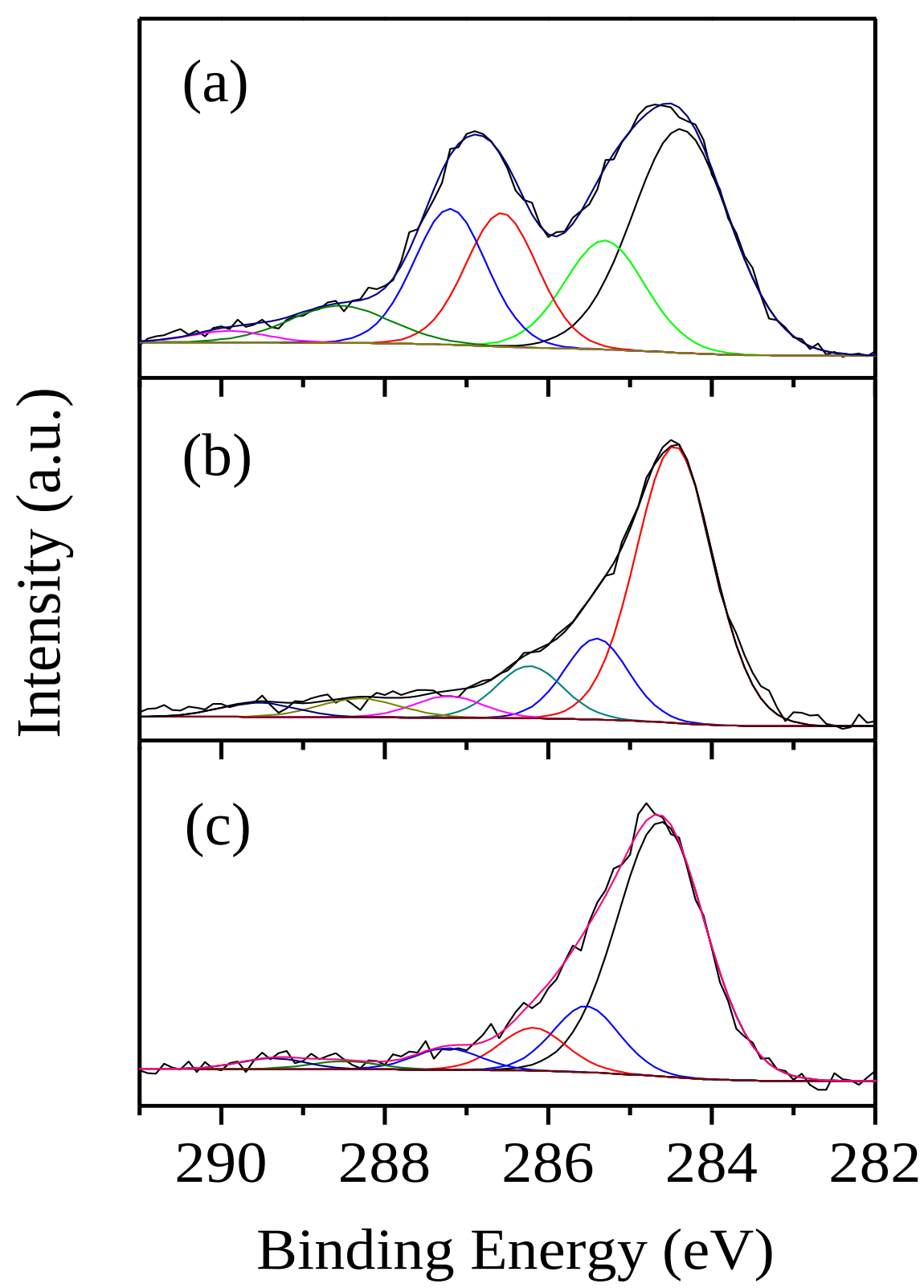

Figure 5. De-convoluted C-1s core level XPS spectra of (a) as-transferred GO sheets and (b) after heat treatment at 400 °C, (c) after heat treatment of AA/GO/AA structure at 200 °C.

Significant increase of $sp^2$-C content, decrease of O/C ratio and X value after the heat treatment (200 °C) of AA/GO/AA structures clearly indicates a substantial reduction of GO in single step reduction process. Reduction of oxygen content is associated with disappearance/decrease of contributions of oxygen functional groups, and the recovery of graphitic network is attributed to the presence carbonaceous products formed by the decomposition of AA molecules. Fatty acid molecules are long chain carbon compounds allied with carbon-carbon double bonds (as shown in Fig. S1), which make these compounds highly reactive in the presence of oxygen/oxygen groups [48]. Further, GO sheets can be effectively reduced by heat treatment in close proximity to AA layers, which may also act as capping/reducing agents, like the other phyto-constituents such as ketones, terpenoids, fatty acids, esters, flavonoids and palm oil leaves extract [37, 49]. However, the SEM and AFM studies carried out on AA/GO/AA structures evidence nearly unaltered morphology of GO sheets before and after heat treatments. On the other hand, near removal of vibrational features allied with AA molecules, relative enhancement of band due to in plane vibrations of $sp^2$ carbons of GO, substantial enhancement of O/C ratio and reduction of I(D)/I(G) ratio together substantiate the reduction of GO sheets during their heat treatment in close proximity of AA molecules. It may also be noted that the extent of GO reduction obtained by heat treatment of sandwich structure at low temperature of 200 °C is significantly superior to the case of GO sheets (without the AA layer) heat treated at 400 °C. These results, along with Raman and FT-IR studies reveal that, GO sheets can be effectively reduced by heat treatment at moderate temperature of 200 °C in close proximity of AA layers and, the extent reduction is similar to those reported via chemical reduction with hydrazine, followed by heat treatment at temperatures in the range of 400–1000 °C [14,15,50].

| Sample (Heat treatment Temperature) | $sp^2$-C | $sp^3$-C | C-O | C=O | COOH | $\pi$-$\pi$* | O/C ratio | X |
|---|---|---|---|---|---|---|---|---|
| GO | 284.4 (42) | 285.3 (15) | 286.6 (17) | 287.2 (17) | 288.6 (6) | 289.9 (3) | 0.46 | 1.16 |
| GO (400 °C) | 284.5 (63) | 285.4 (15) | 286.2 (10) | 287.1 (4) | 288.4 (5) | 289.5 (3) | 0.24 | 0.52 |
| AA/GO/AA (200 °C) | 284.4 (69) | 285.4 (14) | 286.2 (9) | 287.2 (4) | 288.5 (2) | 289.5 (2) | 0.17 | 0.41 |

Table 2. Peak positions (in eV) of various de-convoluted components of C-1*s* core level XPS spectra of GO, AA/GO/AA structure after heat treatment, as indicated. The integrated intensity values (%) are given in parenthesis. The last column gives the values of O/C ratio and X (ratio of non-graphitic carbon to graphitic carbon).

Figs. 6(a-c) show the valence band (VB) spectra of RGO sheets on Si substrate obtained by heat treatment of AA/GO/AA sandwich structure at 200 °C among which, Fig. 6(a) shows the He-II photoelectron spectrum of RGO sheets. The corresponding de-convoluted spectrum is shown in Fig. 6(b) and Table 3 lists the peak positions of de-convoluted components of He-II VB spectrum of RGO sheets obtained through this procedure. The de-convoluted spectrum shows C 2*p*-π peak (3.7-4.0 eV) with considerable intensity, accompanied by the presence of a prominent C 2*p*-(π-σ) band at 5.7-6.0 eV. Apart from these, the C 2*p*-σ band (~6.9 eV), C 2*s*-2*p* mixed band (~9.1 eV) and C 2*s* band (~13.7 eV) were found to be shifted to higher binding energies, as compared to the values reported for GO sheets [51]. Further, an additional peak has been observed at ~11.5 eV, which may be attributed to the C–H 2*p*-σ states [52]. Appearance of C–H 2*p*-σ states is due to the presence of residual AA molecules along with RGO sheets, as indicated by AFM studies. Fig. 6(c) shows the magnified regions near the Fermi level for RGO sheets, where a noticeable increase in density of states (DOS) near the Fermi level has been observed. These features are attributed to increase in intensity of C 2*p*-π states and, proximity of the Fermi level with the valence band edge (VBE) is an indicative of p-type nature of formed RGO [26,51]. In continuation to these studies, Fig. 6(d) shows the secondary electron threshold region of the He-I spectra of RGO sheets obtained after heat treatment of AA sandwich structure. The work function is found to be ~4.0 eV, similar to that observed for chemically reduced GO sheets [51]. The decrease in work function after reduction is attributed to the removal of oxygen functional groups and increase in the graphitic carbon content, consistent with XPS, FT-IR and Raman results.

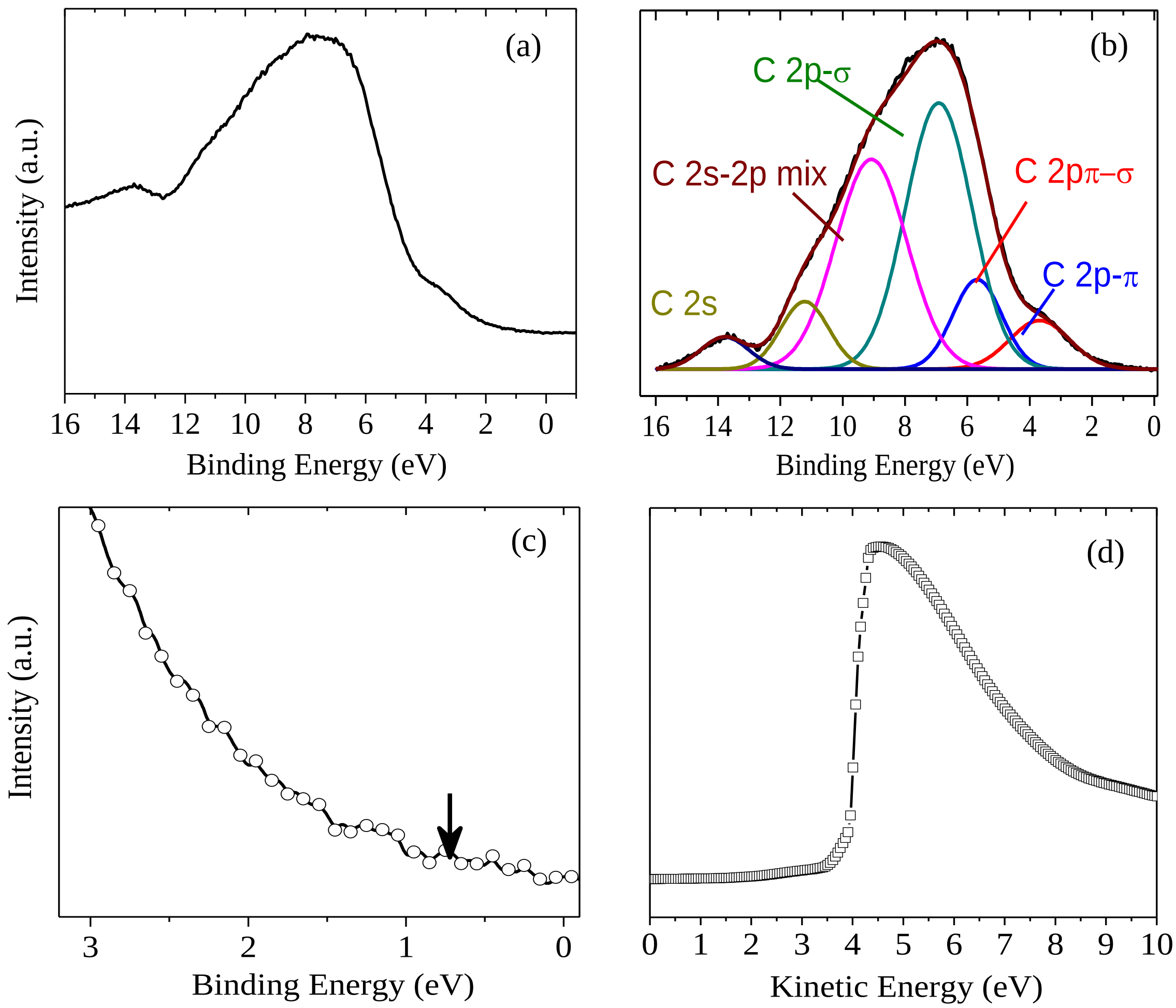

Figure 6. (a) He-II valence band spectrum of AA/GO/AA structures after heat treatment at 200 °C. (b) shows the de-convoluted He-II VB spectra, after background correction and, the magnified regions near Fermi level region are shown in (c). (d) Secondary electron threshold region of He-I spectra of AA/GO/AA structure after heat treatment at 200 °C.

| Sample (Heat treatment temperature) | C 2*p*-π | C 2*p*-(π-σ) | C 2*p*-σ | C 2*s*-2*p* mix | C-H 2*p*-σ | C 2s |
|---|---|---|---|---|---|---|
| AA/GO/AA (200 °C) | 3.7 | 5.7 | 6.9 | 9.1 | 11.2 | 13.7 |

Table 3. De-convoluted peak positions (in eV) of He-II UPS valence band spectra of AA/GO/AA structures after a heat treatment (as indicated).

EFM measurements have been performed to investigate the surface electrical properties of RGO sheets obtained by heat treatment of AA/GO/AA sandwich structure at 200 °C and Fig. 7 shows the AFM and EFM images of obtained RGO sheets. The EFM images correspond to a tip-sample surface separation ($h$) of 20 nm and tip bias voltage ($V_{tip}$) of 2 V. For the purpose of comparison, AFM and EFM images of GO sheets (without the AA layer) on $SiO_2$/Si (Fig. 7(a) and 7(d)) and those obtained after heat treatment at 400 °C in vacuum (Fig. 7(b) and 7(e)) are shown as Fig. 7. The phase image of RGO sheets obtained by heat treatment at 400 °C without the AA layer (Fig. 7(e)) shows a relatively small contrast $\Delta\Phi$ (~ 1.2°). In comparison, the RGO sheets obtained from AA/GO/AA sandwich structure after heat treatment at 200 °C (Fig. 7(f)) showed much larger values of phase contrast for which, $\Delta\Phi$ is in the range of 2.5-2.7 ° at the edges of the sheets. The dependence of $\Delta\Phi$ on $V_{tip}$ in the range of -2 V to +2 V was studied at a constant value of $h$=20 nm and these results are shown in Fig. 8. The data exhibits characteristic parabolic dependence of the phase shift on tip bias voltage, represented by equation (1) in all the cases and are fitted accordingly [53–55].

$$\Delta\Phi = (-Q/2k)\ \Delta C^{ll}(h)(V_{tip} - V_s)^2 \tag{1}$$

where, $\Delta C^{ll}(h)$ is the difference between the second derivatives of the tip-sample capacitance and the tip-bare substrate capacitance, Q and k are respectively, the quality factor and the force constant of the cantilever, $V_{tip}$ is the tip bias voltage and $V_s$ is the local surface potential.

Parabolic dependence is seen at relatively smaller curvature for RGO sheets obtained by heat treatment at 400 °C, without the AA layer. The curvature is substantially larger for the RGO sheets obtained by heat treatment of AA/GO/AA structure at 200 °C. It may be pointed out that for these RGO sheets, the vertex point (the maximum of the parabolic fitting which represents the surface potential of the samples) is in the range of 25–35 mV, indicating their p-type nature.

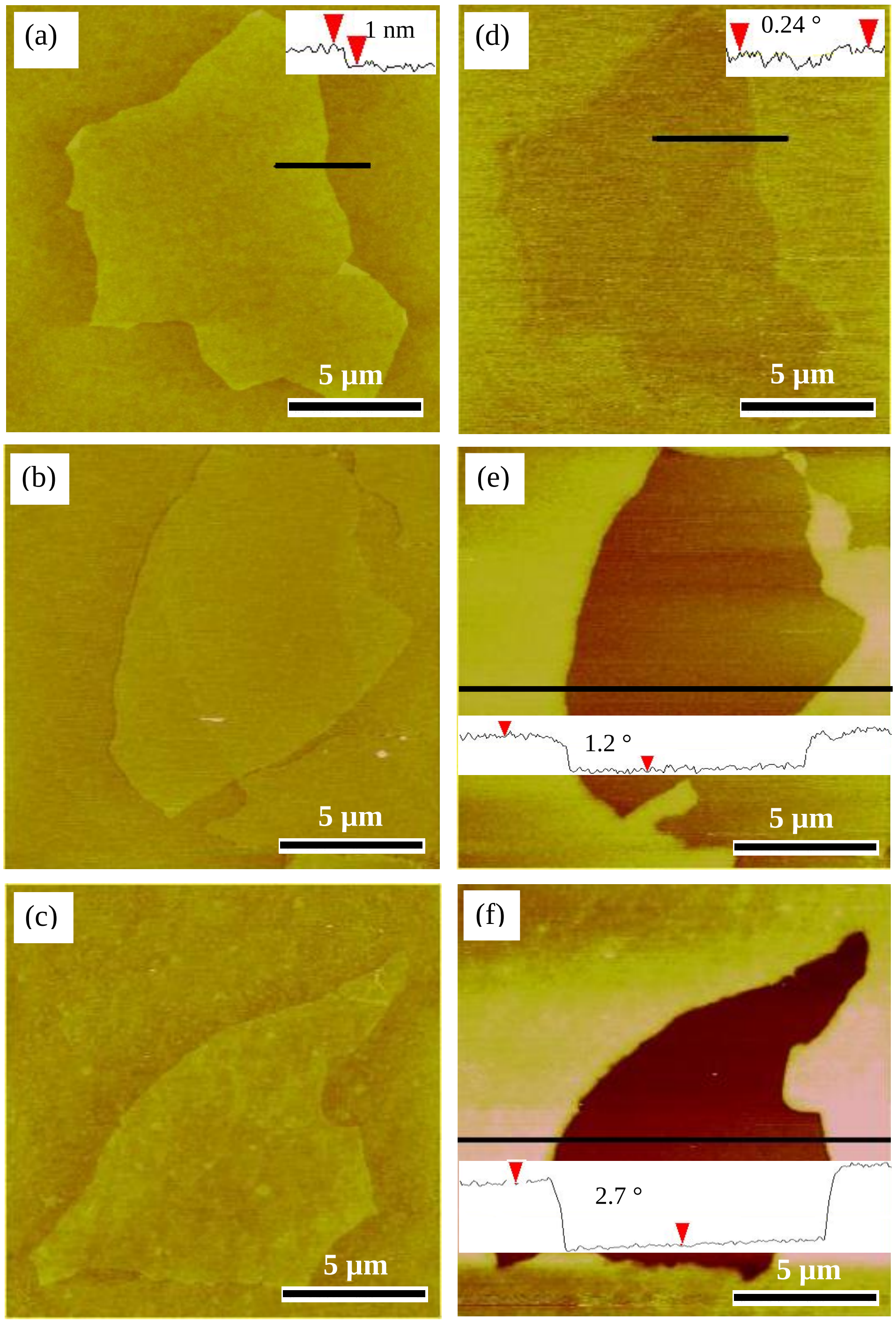

Figure 7. Typical AFM and EFM images of (a,d) GO sheets on $SiO_2$ substrate, (b,e) GO sheets heat treated at 400 °C in vacuum, (c,f) after heat treatment of AA/GO/AA structure at 200 °C. The phase contrast profiles are shown in the EFM images (d-f).

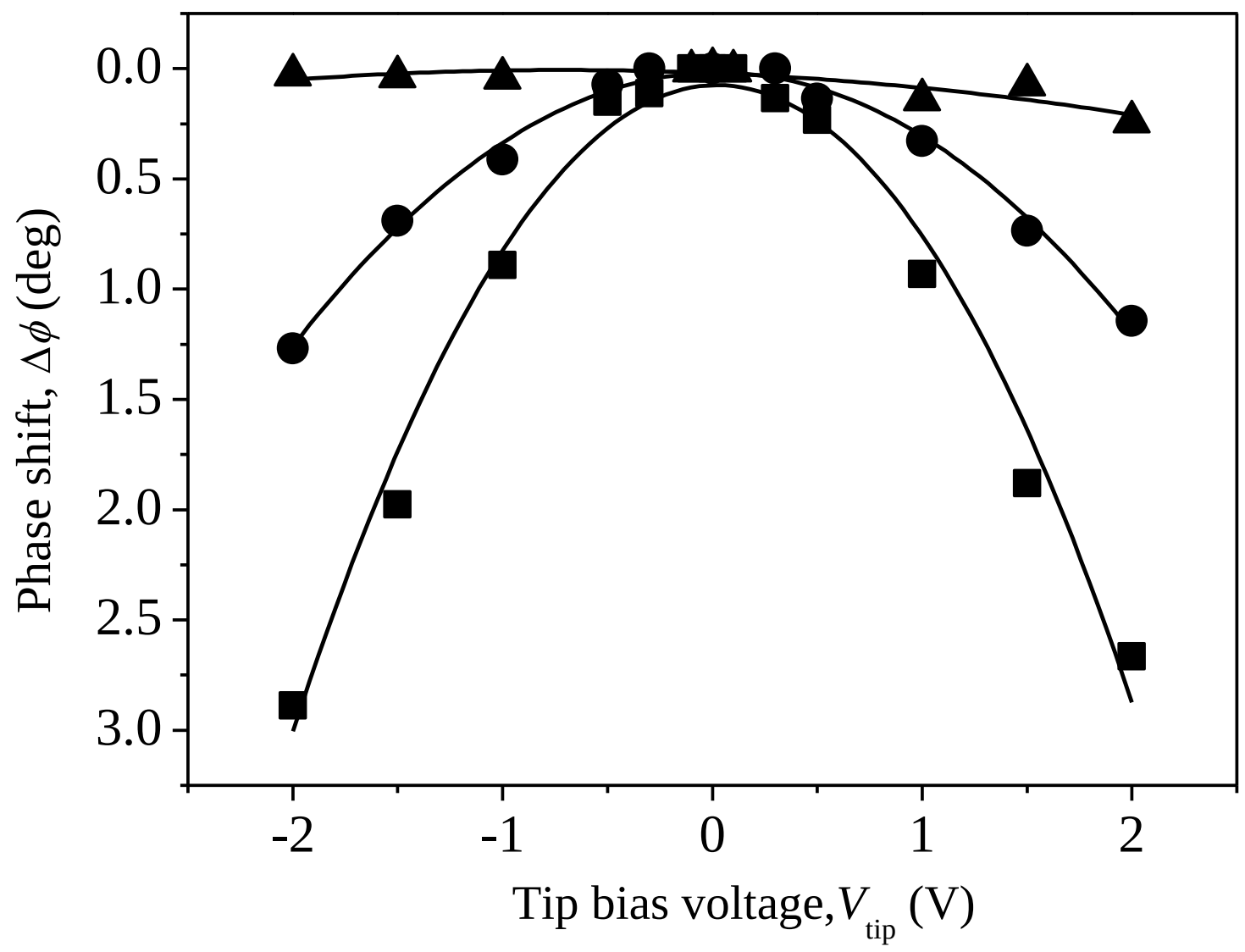

Figure 8. Typical parabolic dependence of $\Delta\Phi$ on tip bias voltage for $h$ = 20 nm at a typical location near the edges of GO (-▲-), RGO obtained by heat treatment at 400 °C in vacuum without AA layer (-●-) and heat treatment of AA/GO/AA structure at 200 °C (-■-) sheets.

The larger curvature of the parabola is an indicative of the higher conductivity of RGO sheets, which is consistent with the corresponding XPS and Raman results. Increase in curvature of parabola with the extent of reduction of GO sheets (Fig. 8) is thus attributed to increase in their dielectric constant, which accompanies the increase in free charge carrier density and conductivity owing to the reduction process, as reported earlier [56,57]. The positive value of vertex point of the parabolic dependence in this case is also indicative of the p-type nature of RGO sheets, consistent with the UPS studies. Further, the phase contrasts (Fig. 8) seen for RGO sheets obtained by heat treatment of AA/GO/AA structures are comparable to those of RGO sheets obtained by chemical reduction with hydrazine, followed by heat treatment in argon at 400 °C for different durations (3 hr (RGO-1) and 6 hr (RGO-2), Fig. S4). It is also noteworthy that the non-destructive EFM technique can be effectively utilized for a qualitative comparison of RGO sheets of varying conductivities.

In order to investigate the effect of de-oxygenation and restoration of graphitic network on the electrical transport properties of RGO sheets, bottom gated field effect transistors (FETs) have been fabricated on isolated RGO monolayer sheets on $SiO_2$/Si substrates, obtained by heat treatment of AA/GO/AA sandwich structure. These characterizations were carried on multiple samples, each having 20 to 25 devices in a single chip. The channel length and width were maintained in the range of 12–25 µm and 10–30 µm, respectively. The SEM images revealed that the RGO sheets had a variety of irregular shapes and sizes. As a result, the device channel's length and width were measured by taking into account the average of both sides, respectively. Fig. 9 shows the typical $I_{DS}$-$V_{DS}$ characteristics of RGO sheets measured in bottom gated-FET geometry. For better comparison, the $I_{DS}$-$V_{DS}$ characteristics of RGO sheets obtained by heat treatment of GO sheets (without AA layers) in vacuum at 400 °C were also examined along with those obtained by heat treatment of AA/GO/AA structures at 200 °C. It may be noted that, these $I_{DS}$-$V_{DS}$ characteristics were recorded over multiple RGO sheets on different FET devices prepared under similar conditions and typical results are presented in Fig. 9. The RGO sheets obtained by heat treatment at 400 °C exhibited non-linear I-V curves and their conductivities were observed in the range of $10^{-2}$ – $10^{-1}$ S/cm, which are much larger than those of typical GO sheets [15]. In contrast, the RGO sheets obtained by heat treatment of AA/GO/AA structures exhibited linear I-V curves and displayed substantially higher conductivities in the range of 2–7 S/cm. A typical top view SEM image of an RGO monolayer-based FET geometry is shown in the inset of Fig. 9(b).

Typical transfer characteristics ($I_{DS}$-$V_{GS}$ curve at $V_{DS}$ = 1 V) of back gated FETs based on RGO sheets obtained by heat treatment at 400 °C (without AA layer) as well as RGO sheets obtained by heat treatment of AA/GO/AA structures at 200 °C were measured in the gate voltage range of -20 V to +20 V and the obtained results are shown in Fig. 9. For the FETs fabricated with RGO sheets obtained by heat treatment at 400 °C, $I_{DS}$ decreases with decrease of negative gate voltage and continues to decrease with raise of positive gate voltage to +20 V, thus displaying a p-type behaviour. On the other hand, the FETs fabricated with RGO sheets obtained via heat treatment of AA/GO/AA structure displayed larger currents of 1-2 order higher magnitude and p-type behavior. This may be attributed to the influence of surface adsorbed of oxygen molecules or residual hydroxyl groups on RGO sheets, which can serve as deep electron traps and suppress n-type conductivity [15,58]. For the RGO sheets heat treated at 400 °C (without the AA layer), the hole mobility was found to be in the range of 0.001–0.04 $cm^2$/V-s. In comparison, for the RGO sheets obtained by the heat treatment of AA/GO/AA structure, the hole mobilities were observed to be in the range of 0.03–2 $cm^2$/V-s. Most importantly, the conductivity and field effect mobility values of RGO sheets obtained by low temperature heat treatment of AA/GO/AA sandwich structures are comparable (Table 4) to those RGO monolayers obtained by various earlier reported chemical/thermal reduction methods [15,59].

| Reducing agent & method of fabrication | Process of reduction | C/O | I(D)/I(G) | σ (S/cm) or $R_s$ (kΩ/□) | Type of layers or thickness of RGO (nm) | References |
|---|---|---|---|---|---|---|
| Sulfur containing compounds | Solution | 6-8 | 0.95-1.22 | - | Mono and multilayers | [20] |
| Ammonia plasma, 5-20 min, 10 W | Solid state | 0.038 | 1.9 | 1–80 S/cm | Monolayers | [25] |
| Hydrogen plasma, 30 sec, 50 °C, 15 W | Solid state | 6.25 | 2.3 | (0.2–31) S/cm | Monolayers | [26] |
| Graphite mediated annealing@ 600-1000 °C | Solid state | 0.2 | 1.5-2 | (2-3) x$10^3$ S/cm | Monolayers | [43] |
| Hydrazine vapor reduction @80 °C for 24 hr and/or annealing in 20% $H_2$/Ar 800 -1000 °C for 2 hr, | Solid state | - | 1.16-1.55 | 1005-188 kΩ/□ | Monolayers | [59] |
| $CH_4$/$H_2$ plasma in less than 6 min at 600 °C | Solid state | 10 | 0.8 | 83 kΩ/□ | Monolayers | [60] |
| Hydroiodic acid (HI)/ trifluoroacetic acid (TFA), 80 °C, 10 min + annealing with ethanol at 800 °C in Ar/$H_2$, 30 min | Solid state and solution | | 0.34 | 62.1 × $10^3$S/cm, 0.46kΩ/□ | Monolayers | [61] |
| Polymer-Encapsulated @1000−1100 °CAr | Solid state | 6 | 1.46 | 580 S/cm, 23.8 kΩ/□ | Mono and Bilayers | [62] |
| Arachidic acid @ 1hr at 200 °C | Solid state | 5.88 | 1.1 | ~ 2-7 S/cm, ~1000 kΩ/□ | Monolayer (1 nm) | This work |

Table 4. A comparative summary of various chemical/thermal reduction methods and their effect on the degree of reduction in terms of C/O ratio, I(D)/I(G) ratio, conductivity values and nature/thickness of the RGO sheets.

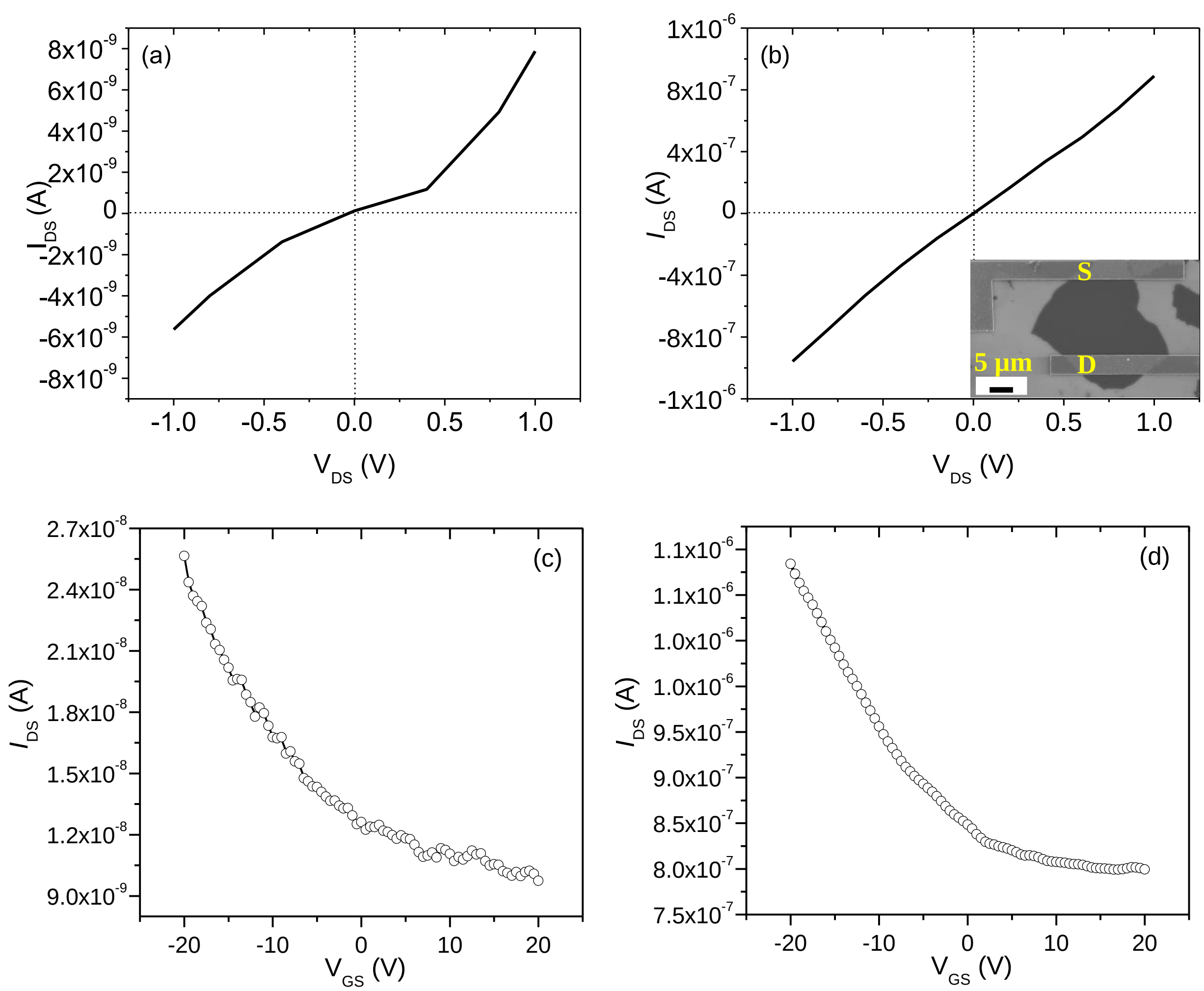

Figure 9. $I_{DS}$-$V_{DS}$ plots of (a) RGO sheets obtained by heat treatment at 400 °C in vacuum without AA layer, and (b) AA/GO/AA structure after heat treatment at 200 °C, along with SEM image of a typical RGO monolayer sheet in two probe contact geometry as inset. Transfer characteristics of the bottom-gated FET employing (c) RGO sheets obtained by heat treatment at 400 °C in vacuum without AA layer, (d) AA/GO/AA structure after heat treatment at 200 °C.

Further, to examine the effect of (a) escalation of heat treatment temperature and (b) multiplying AA layers in sandwich structure on extent of GO reduction, two entirely different reduction procedures were adopted. In the first case, typical as-transferred AA/GO/AA structure was subjected to heat treatment in vacuum (~$10^{-5}$ mbar) at 400 °C for 1 hr. In the second case, a set of three layers of AA have been used instead of AA monolayer in the sandwich structure (3AA/GO/3AA) and subjected to heat treatment in a vacuum (~$10^{-5}$ mbar) at 200 °C for 1 hr. The XPS data for both these types of samples are shown in Table S1. As compared to the optimized samples (AA/GO/AA heat treated at 200 °C), no further or significant changes in reduction have been observed with either escalation of heat treatment temperature or increase of AA layers (Table S1). Interestingly, GO sheets were clearly seen after heat treatment of 3AA/GO/3AA structure at 200 °C, despite the availability of larger quantity AA in precursor structure (Fig. S3). However, the height profile measurements carried out on these samples showed a largely uneven surface with high surface roughness (0.6–0.8 nm), in comparison with the limited roughness seen for the case of AA/GO/AA structure (~0.2 nm, Fig. 2(b)). These features are attributed to incomplete removal of AA molecules from the GO sheet surface. In brief, feasibility of GO reduction via heat treatment of AA/GO/AA structures and the extent of reduction were comprehensively analyzed in the present work in light of XPS, Raman, FT-IR and electrical characterization results, which collectively indicated significant reduction of oxygen content and restoration of graphitic network. Most importantly, the extent GO reduction is equivalent to those reported via chemical reduction with hydrazine, followed by heat treatment at temperatures in the range of 400–1000 °C [14,15,50]. The proposed method has been further verified by escalation of heat treatment temperature and multiplying AA layers (3AA/GO/3AA) in sandwich structure, which reiterated the reproducibility of results, significant reduction of GO and evidence formation of RGO.

## 4. Conclusions

The present work introduces a novel, single-step and environment friendly reduction method via heat treatment of AA/GO/AA sandwich structure developed by sequential LB processing at moderate temperature of 200$^{o}$C. Heat treatment of AA/GO/AA sandwich structure does not lead to significant change in the thickness, surface density and overall morphology of the GO sheets, except the presence of residual AA molecules. However, heat treatment of AA/GO/AA sandwich structure at 200 °C causes substantial reduction of GO, resulting in formation of RGO layers with $sp^2$-C content of ~69%, O/C ratio of ~0.17, non-graphitic/graphitic carbon ratio of ~0.4 and I(D)/I(G) ratio of ~1.1. The extent of reduction and electrical properties of RGO sheets obtained by heat treatment of AA/GO/AA sandwich structures at low temperatures (200°C) were further analyzed with the aid of valence band electronic structure studies using UPS, non-destructive assessment of conductivity by EFM measurements and electrical characterization studies with bottom gated FETs. Precisely, the RGO sheets obtained by heat treatment of AA/GO/AA sandwich structure exhibited a dominantly p-type conductivity in the range of 2–7 S/cm and hole mobility in the range of 0.03–2 $cm^2$/V-s. These electrical parameters are comparable with those reported for RGO sheets usually obtained by hydrazine vapour treatment, followed by heat treatment in the range of 400–1000 °C. Thus, GO sheets can be effectively reduced by heat treatment of LB-transferred AA/GO/AA sandwich structures, where the presence of AA molecules in close proximity of GO sheets play a significant role in reduction of GO sheets and substantial recovery of graphitic network. The solid state reduction procedure proposed in the present work not only offers a simple, easily adoptable and environment friendly route for the reduction of GO sheets, but also facilitates their applications in development of solid state device structures.

**Acknowledgments**

Prof. S.S. Major of Department of Physics, IIT Bombay is gratefully acknowledged for all the support and insightful discussions. Authors thankfully acknowledges FIST (PHYSICS)-IRCC Central SPM Facility, Centre for Excellence in Nanoelectronics, Centre for Research in Nanotechnology and Science, IIT Bombay for SPM, SEM, XPS, Raman spectroscopy and electrical measurements. This work is supported through ANTARES project that has received funding from the European Union's Horizon 2020 research and innovation programme under grant agreement SGA-CSA. No. 739570 under FPA No. 664387.

**References:**

[1] G. Eda, M. Chhowalla, Chemically derived graphene oxide: Towards large-area thin-film electronics and optoelectronics, Advanced Materials 22 (2010) 2392–2415. https://doi.org/10.1002/adma.200903689.

[2] Y. Hu, H. Gao, Chemical synthesis of reduced graphene oxide: a review, Minerals and Mineral Materials 2 (2023). https://doi.org/10.20517/mmm.2023.07.

[3] Y. Jin, Y. Zheng, S.G. Podkolzin, W. Lee, Band gap of reduced graphene oxide tuned by controlling functional groups, J Mater Chem C Mater 8 (2020) 4885–4894. https://doi.org/10.1039/c9tc07063j.

[4] O.C. Compton, S.T. Nguyen, Graphene oxide, highly reduced graphene oxide, and graphene: Versatile building blocks for carbon-based materials, Small 6 (2010) 711–723. https://doi.org/10.1002/smll.200901934.

[5] C.K. Chua, M. Pumera, Chemical reduction of graphene oxide: A synthetic chemistry viewpoint, Chem Soc Rev 43 (2014) 291–312. https://doi.org/10.1039/c3cs60303b.

[6] N. Sharma, M. Arif, S. Monga, M. Shkir, Y.K. Mishra, A. Singh, Investigation of bandgap alteration in graphene oxide with different reduction routes, Appl Surf Sci 513 (2020). https://doi.org/10.1016/j.apsusc.2020.145396.

[7] Z. Bo, X. Shuai, S. Mao, H. Yang, J. Qian, J. Chen, J. Yan, K. Cen, Green preparation of reduced graphene oxide for sensing and energy storage applications, Sci Rep 4 (2014). https://doi.org/10.1038/srep04684.

[8] M.J. Allen, V.C. Tung, R.B. Kaner, Honeycomb carbon: A review of graphene, Chem Rev 110 (2010) 132–145. https://doi.org/10.1021/cr900070d.

[9] H. Pant, S. Petnikota, V.S.S.S. Vadali, Review—Brief Review of the Solid-State Graphenothermal Reduction for Processing Metal Oxide-Reduced Graphene Oxide Nanocomposites for Energy Applications, ECS Journal of Solid State Science and Technology 10 (2021) 031002. https://doi.org/10.1149/2162-8777/abe8b4.

[10] X. Lv, Y. Huang, Z. Liu, J. Tian, Y. Wang, Y. Ma, J. Liang, S. Fu, X. Wan, Y. Chen, Photoconductivity of bulk-film-based graphene sheets, Small 5 (2009) 1682–1687. https://doi.org/10.1002/smll.200900044.

[11] X. Wang, L. Zhi, K. Müllen, Transparent, conductive graphene electrodes for dye-sensitized solar cells, Nano Lett 8 (2008) 323–327. https://doi.org/10.1021/nl072838r.

[12] S. Gilje, S. Han, M. Wang, K.L. Wang, R.B. Kaner, A chemical route to graphene for device applications, Nano Lett 7 (2007) 3394–3398. https://doi.org/10.1021/nl0717715.

[13] V. Lee, L. Whittaker, C. Jaye, K.M. Baroudi, D.A. Fischer, S. Banerjee, Large-area chemically medified graphene films: Electrophoretic deposition and characterization by soft X-ray absorption spectroscopy, Chemistry of Materials 21 (2009) 3905–3916. https://doi.org/10.1021/cm901554p.

[14] D.S. Sutar, P.K. Narayanam, G. Singh, V.D. Botcha, S.S. Talwar, R.S. Srinivasa, S.S. Major, Spectroscopic studies of large sheets of graphene oxide and reduced graphene oxide monolayers prepared by Langmuir-Blodgett technique, Thin Solid Films 520 (2012). https://doi.org/10.1016/j.tsf.2012.05.018.

[15] V.D. Botcha, G. Singh, P.K. Narayanam, S.S. Talwar, R.S. Srinivasa, S.S. Major, A "modified" Langmuir-Blodgett technique for transfer of graphene oxide monolayer sheets on solid substrates, Mater Res Express 3 (2016). https://doi.org/10.1088/2053-1591/3/3/035002.

[16] S. Stankovich, D.A. Dikin, R.D. Piner, K.A. Kohlhaas, A. Kleinhammes, Y. Jia, Y. Wu, S.B.T. Nguyen, R.S. Ruoff, Synthesis of graphene-based nanosheets via chemical reduction of exfoliated graphite oxide, Carbon N Y 45 (2007) 1558–1565. https://doi.org/10.1016/j.carbon.2007.02.034.

[17] H.J. Shin, K.K. Kim, A. Benayad, S.M. Yoon, H.K. Park, I.S. Jung, M.H. Jin, H.K. Jeong, J.M. Kim, J.Y. Choi, Y.H. Lee, Efficient reduction of graphite oxide by sodium borohydride and its effect on electrical conductance, Adv Funct Mater 19 (2009) 1987–1992. https://doi.org/10.1002/adfm.200900167.

[18] X. Fan, W. Peng, Y. Li, X. Li, S. Wang, G. Zhang, F. Zhang, Deoxygenation of exfoliated graphite oxide under alkaline conditions: a green route to graphene preparation, Advanced Materials 20 (2008) 4490–4493. https://doi.org/10.1002/adma.200801306.

[19] S. Pei, J. Zhao, J. Du, W. Ren, H.M. Cheng, Direct reduction of graphene oxide films into highly conductive and flexible graphene films by hydrohalic acids, Carbon N Y 48 (2010) 4466–4474. https://doi.org/10.1016/j.carbon.2010.08.006.

[20] W. Chen, L. Yan, P.R. Bangal, Chemical reduction of graphene oxide to graphene by sulfur-containing compounds, Journal of Physical Chemistry C 114 (2010) 19885–19890. https://doi.org/10.1021/jp107131v.

[21] D.R. Dreyer, S. Murali, Y. Zhu, R.S. Ruoff, C.W. Bielawski, Reduction of graphite oxide using alcohols, J Mater Chem 21 (2011) 3443–3447. https://doi.org/10.1039/c0jm02704a.

[22] A. Furst, R.C. Berlo, S. Hooton, Chem. Rev. 65 (1965) 51–68. https://doi.org/10.1021/cr60233a002.

[23] A.E.D. Mahmoud, Eco-friendly reduction of graphene oxide via agricultural byproducts or aquatic macrophytes, Mater Chem Phys 253 (2020). https://doi.org/10.1016/j.matchemphys.2020.123336.

[24] F. Catania, E. Marras, M. Giorcelli, P. Jagdale, L. Lavagna, A. Tagliaferro, M. Bartoli, A review on recent advancements of graphene and graphene-related materials in biological applications, Applied Sciences (Switzerland) 11 (2021) 1–21. https://doi.org/10.3390/app11020614.

[25] G. Singh, D.S. Sutar, V. Divakar Botcha, P.K. Narayanam, S.S. Talwar, R.S. Srinivasa, S.S. Major, Study of simultaneous reduction and nitrogen doping of graphene oxide Langmuir-Blodgett monolayer sheets by ammonia plasma treatment, Nanotechnology 24 (2013). https://doi.org/10.1088/0957-4484/24/35/355704.

[26] G. Singh, V.D. Botcha, D.S. Sutar, P.K. Narayanam, S.S. Talwar, R.S. Srinivasa, S.S. Major, Near room temperature reduction of graphene oxide Langmuir-Blodgett monolayers by hydrogen plasma, Physical Chemistry Chemical Physics 16 (2014). https://doi.org/10.1039/c4cp00875h.

[27] M. Zhou, Y. Wang, Y. Zhai, J. Zhai, W. Ren, F. Wang, S. Dong, Controlled synthesis of large-area and patterned electrochemically reduced graphene oxide films, Chemistry - A European Journal 15 (2009) 6116–6120. https://doi.org/10.1002/chem.200900596.

[28] J. Zhang, H. Yang, G. Shen, P. Cheng, J. Zhang, S. Guo, Reduction of graphene oxide vial-ascorbic acid, Chemical Communications 46 (2010) 1112–1114. https://doi.org/10.1039/b917705a.

[29] R. Joshi, A. De Adhikari, A. Dey, I. Lahiri, Green reduction of graphene oxide as a substitute of acidic reducing agents for supercapacitor applications, Materials Science and Engineering: B 287 (2023). https://doi.org/10.1016/j.mseb.2022.116128.

[30] S. Sadhukhan, A. Bhattacharyya, D. Rana, T.K. Ghosh, J.T. Orasugh, S. Khatua, K. Acharya, D. Chattopadhyay, Synthesis of RGO/NiO nanocomposites adopting a green approach and its photocatalytic and antibacterial properties, Mater Chem Phys 247 (2020). https://doi.org/10.1016/j.matchemphys.2020.122906.

[31] M. Nasrollahzadeh, S. Mohammad Sajadi, A. Rostami-Vartooni, M. Alizadeh, M. Bagherzadeh, Green synthesis of the Pd nanoparticles supported on reduced graphene oxide using barberry fruit extract and its application as a recyclable and heterogeneous

catalyst for the reduction of nitroarenes, J Colloid Interface Sci 466 (2016) 360–368. https://doi.org/10.1016/j.jcis.2015.12.036.

[32] E. Vatandost, A. Ghorbani-HasanSaraei, F. Chekin, S. NaghizadehRaeisi, S.A. Shahidi, Green tea extract assisted green synthesis of reduced graphene oxide: Application for highly sensitive electrochemical detection of sunset yellow in food products, Food Chem X 6 (2020). https://doi.org/10.1016/j.fochx.2020.100085.

[33] J. Wu, Y. Wei, H. Ding, Z. Wu, X. Yang, Z. Li, W. Huang, X. Xie, K. Tao, X. Wang, Green Synthesis of 3D Chemically Functionalized Graphene Hydrogel for High-Performance NH3 and NO2 Detection at Room Temperature, ACS Appl Mater Interfaces 12 (2020) 20623–20632. https://doi.org/10.1021/acsami.0c00578.

[34] F. Yousefimehr, S. Jafarirad, R. Salehi, M.S. Zakerhamidi, Facile fabricating of rGO and Au/rGO nanocomposites using Brassica oleracea var. gongylodes biomass for non-invasive approach in cancer therapy, Sci Rep 11 (2021). https://doi.org/10.1038/s41598-021-91352-7.

[35] A.E.D. Mahmoud, N. El-Maghrabi, M. Hosny, M. Fawzy, Biogenic synthesis of reduced graphene oxide from Ziziphus spina-christi (Christ's thorn jujube) extracts for catalytic, antimicrobial, and antioxidant potentialities, Environmental Science and Pollution Research 29 (2022) 89772–89787. https://doi.org/10.1007/s11356-022-21871-x.

[36] U. Chasanah, W. Trisunaryanti, H.S. Oktaviano, Triyono, I. Santoso, D.A. Fatmawati, Study of green reductant effects of highly reduced graphene oxide production and their characteristics, Communications in Science and Technology 7 (2022) 103–111. https://doi.org/10.21924/cst.7.2.2022.906.

[37] M.S. Amir Faiz, C.A. Che Azurahanim, S.A. Raba'ah, M.Z. Ruzniza, Low cost and green approach in the reduction of graphene oxide (GO) using palm oil leaves extract for potential in industrial applications, Results Phys 16 (2020). https://doi.org/10.1016/j.rinp.2020.102954.

[38] B. Meka Chufa, B. Abdisa Gonfa, T. Yohannes Anshebo, G. Adam Workneh, A Novel and Simplest Green Synthesis Method of Reduced Graphene Oxide Using Methanol Extracted Vernonia Amygdalina: Large-Scale Production, Advances in Condensed Matter Physics 2021 (2021). https://doi.org/10.1155/2021/6681710.

[39] E.W. Omagamre, J. Fotouhi, T. Washington, C. Ahuchaogu, Y. Mansurian, S. Kundu, K.S. Das, Molecular filtration by reduced graphene oxide sandwiched between wood sheets, arXiv:2209.12983. https://doi.org/10.48550/arXiv.2209.12983.

[40] A. Longo, M. Palomba, G. Carotenuto, Green solid-state chemical reduction of graphene oxide supported on a paper substrate, Coatings 10 (2020). https://doi.org/10.3390/coatings10070693.

[41] A. Ulman (ed.), Organic thin films and surfaces: Directions for the nineties. Thin Films, vol. 21, Academic Press, Boston (1995).

[42] V.D. Botcha, P.K. Narayanam, G. Singh, S.S. Talwar, R.S. Srinivasa, S.S. Major, Effect of substrate and subphase conditions on the surface morphology of graphene oxide sheets prepared by Langmuir-Blodgett technique, Colloids Surf APhysicochem Eng Asp 452 (2014). https://doi.org/10.1016/j.colsurfa.2014.03.077.

[43] G. Singh, V. Divakar Botcha, D.S. Sutar, S.S. Talwar, R.S. Srinivasa, S.S. Major, Graphite mediated reduction of graphene oxide monolayer sheets, Carbon N Y 95 (2015). https://doi.org/10.1016/j.carbon.2015.08.067.

[44] J.F. Rabolt, F.C. Burns, N.E. Schlotter, J.D. Swalen, Anisotropic orientation in molecular monolayers by infrared spectroscopy, J Chem Phys 78 (1982) 946–952. https://doi.org/10.1063/1.444799.

[45] M. Acik, G. Lee, C. Mattevi, M. Chhowalla, K. Cho, Y.J. Chabal, Unusual infrared-absorption mechanism in thermally reduced graphene oxide, Nat Mater 9 (2010) 840–845. https://doi.org/10.1038/nmat2858.

[46] S. Stankovich, R.D. Piner, X. Chen, N. Wu, S.T. Nguyen, R.S. Ruoff, Stable aqueous dispersions of graphitic nanoplatelets via the reduction of exfoliated graphite oxide in the presence of poly(sodium 4-styrenesulfonate), J Mater Chem 16 (2006) 155–158. https://doi.org/10.1039/b512799h.

[47] A.C. Ferrari, J. Robertson, Interpretation of Raman spectra of disordered and amorphous carbon, Phys Rev B 61 (2000) 14095–14107. https://doi.org/10.1103/PhysRevB.61.14095.

[48] A. Mukhametov, D. Dautkanova, A. Kazhymurat, M. Yerbulekova, G. Aitkhozhayeva, The Effects of Heat Treatment on the Oxidation Resistance and Fatty Acid Composition of the Vegetable Oil Blend, J Oleo Sci 72 (2023) 597–604. https://doi.org/10.5650/jos.ess23010.

[49] A.E.D. Mahmoud, N. El-Maghrabi, M. Hosny, M. Fawzy, Biogenic synthesis of reduced graphene oxide from Ziziphus spina-christi (Christ’s thorn jujube) extracts for catalytic, antimicrobial, and antioxidant potentialities, Environmental Science and Pollution Research 29 (2022) 89772–89787. https://doi.org/10.1007/s11356-022-21871-x.

[50] R.M. Rudenko, O.O. Voitsihovska, V.N. Poroshin, Enhancement of electrical conductivity of hydrazine-reduced graphene oxide under thermal annealing in hydrogen atmosphere, Mater Lett 331 (2023). https://doi.org/10.1016/j.matlet.2022.133476.

[51] D.S. Sutar, G. Singh, V. Divakar Botcha, Electronic structure of graphene oxide and reduced graphene oxide monolayers, Appl Phys Lett 101 (2012). https://doi.org/10.1063/1.4749841.

[52] S.C. Ray, K.P.K. Kumar, H.M. Tsai, J.W. Chiou, C.W. Pao, W.F. Pong, M.H. Tsai, B.H. Wu, C.R. Sheu, C.C. Chen, F.C.N. Hong, H.H. Cheng, A. Dalakyan, Studies of ion irradiation effects in hydrogenated amorphous carbon thin films by X-ray absorption and photoemission spectroscopy, Thin Solid Films 516 (2008) 3374–3377. https://doi.org/10.1016/j.tsf.2007.10.020.

[53] C. Staii, A.T. Johnson, N.J. Pinto, Quantitative analysis of scanning conductance microscopy, Nano Lett 4 (2004) 859–862. https://doi.org/10.1021/nl049748w.

[54] D.D. Kulkarni, S. Kim, M. Chyasnavichyus, K. Hu, A.G. Fedorov, V. V. Tsukruk, Chemical reduction of individual graphene oxide sheets as revealed by electrostatic force microscopy, J Am Chem Soc 136 (2014) 6546–6549. https://doi.org/10.1021/ja5005416.

[55] S.S. Datta, D.R. Strachan, E.J. Mele, A.T.C. Johnson, Surface Potentials and layer charge distributions in Few-Layer Graphene films, Nano Lett 9 (2009) 7–11. https://doi.org/10.1021/nl8009044.

[56] C. Gómez-Navarro, F.J. Guzmán-Vázquez, J. Gómez-Herrero, J.J. Saenz, G.M. Sacha, Fast and non-invasive conductivity determination by the dielectric response of reduced graphene oxide: An electrostatic force microscopy study, Nanoscale 4 (2012) 7231–7236. https://doi.org/10.1039/c2nr32640j.

[57] M. Jaafar, G. López-Polín, C. Gómez-Navarro, J. Gómez-Herrero, Step like surface potential on few layered graphene oxide, Appl Phys Lett 101 (2012). https://doi.org/10.1063/1.4773357.

[58] V.C. Tung, M.J. Allen, Y. Yang, R.B. Kaner, High-throughput solution processing of large-scale graphene, Nat Nanotechnol 4 (2009) 25–29. https://doi.org/10.1038/nnano.2008.329.

[59] C.Y. Su, Y. Xu, W. Zhang, J. Zhao, X. Tang, C.H. Tsai, L.J. Li, Electrical and spectroscopic characterizations of ultra-large reduced graphene oxide monolayers, Chemistry of Materials 21 (2009) 5674–5680. https://doi.org/10.1021/cm902182y.

[60] Y. Morikuni, K. Kanishka, P. Viswanath, M. Hara, M. Yoshimura, Rapid and facile fabrication of conducting monolayer reduced graphene oxide films by methane plasma-assisted reduction, Appl Surf Sci 569 (2021). https://doi.org/10.1016/j.apsusc.2021.151022.

[61] K.K.H. De Silva, K. Shibata, P. Viswanath, H.H. Huang, M. Yoshimura, High-Quality Monolayer Reduced Graphene Oxide Films via Combined Chemical Reduction and Ethanol-Assisted Defect Restoration, Adv Mater Interfaces 9 (2022). https://doi.org/10.1002/admi.202200503.

[62] M. Savchak, N. Borodinov, R. Burtovyy, M. Anayee, K. Hu, R. Ma, A. Grant, H. Li, D.B. Cutshall, Y. Wen, G. Koley, W.R. Harrell, G. Chumanov, V. Tsukruk, I. Luzinov, Highly Conductive and Transparent Reduced Graphene Oxide Nanoscale Films via Thermal Conversion of Polymer-Encapsulated Graphene Oxide Sheets, ACS Appl Mater Interfaces 10 (2018) 3975–3985. https://doi.org/10.1021/acsami.7b16500.

**Electronic Supplementary Information**

# Green and Solid State Reduction of GO Monolayers Sandwiched between Arachidic Acid LB Layers

V. Divakar Botcha[a,*], Pavan K. Narayanam[b,c]

[a]BioSense Institute, University of Novi Sad, Novi Sad, 21000 Serbia

[b]Homi Bhabha National Institute, Training School Complex, Anushaktinagar, Mumbai 400094, India

[c]Materials Chemistry & Metal Fuel Cycle Group, Indira Gandhi Centre for Atomic Research, Kalpakkam 603102, India

*Corresponding author email: divakarbotcha@gmail.com

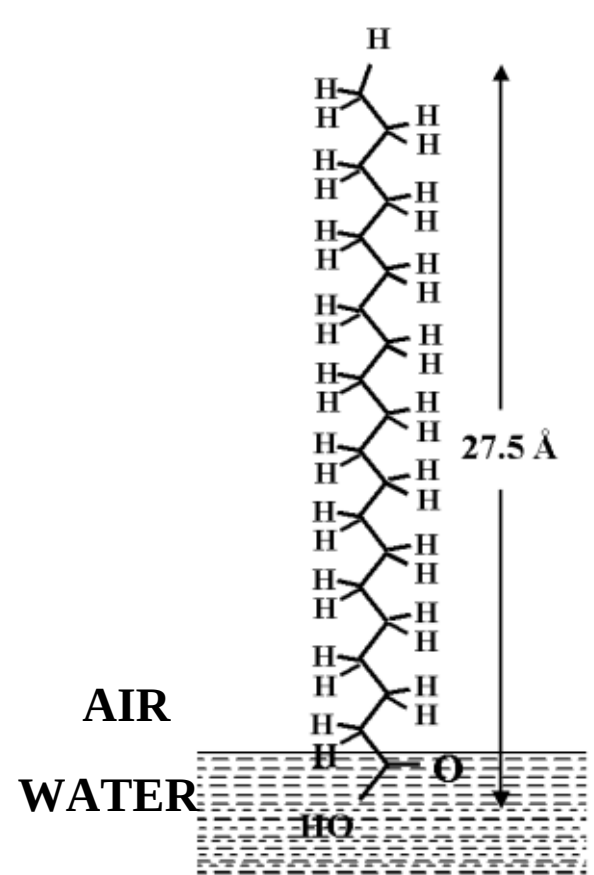

Figure S1. Schematic diagram of an arachidic acid (AA) molecule at the air-water interface.

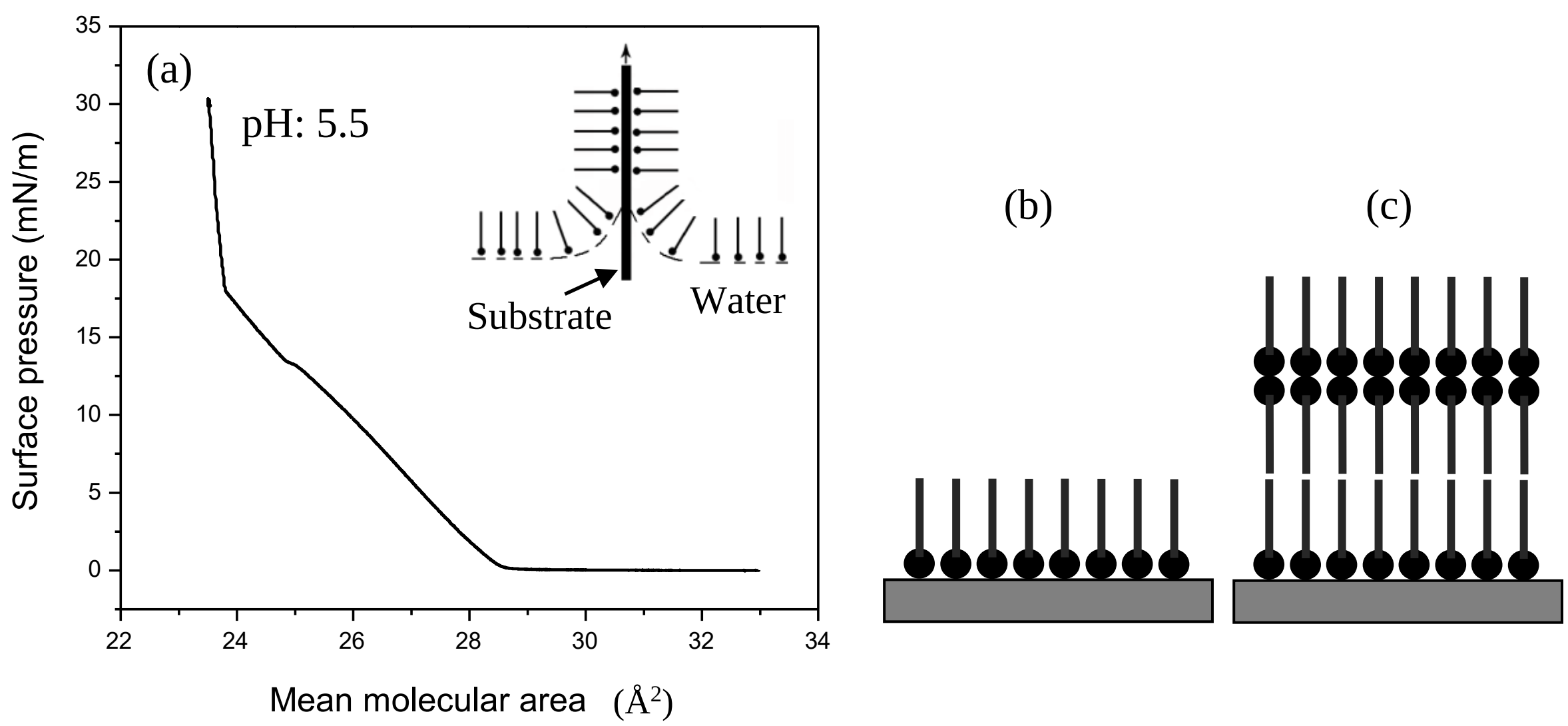

Figure S2. (a) Typical π-A isotherm for an AA monolayer at subphase pH of 5.5 and the schematics of (b) one monolayer of AA and (c) three monolayers of AA.

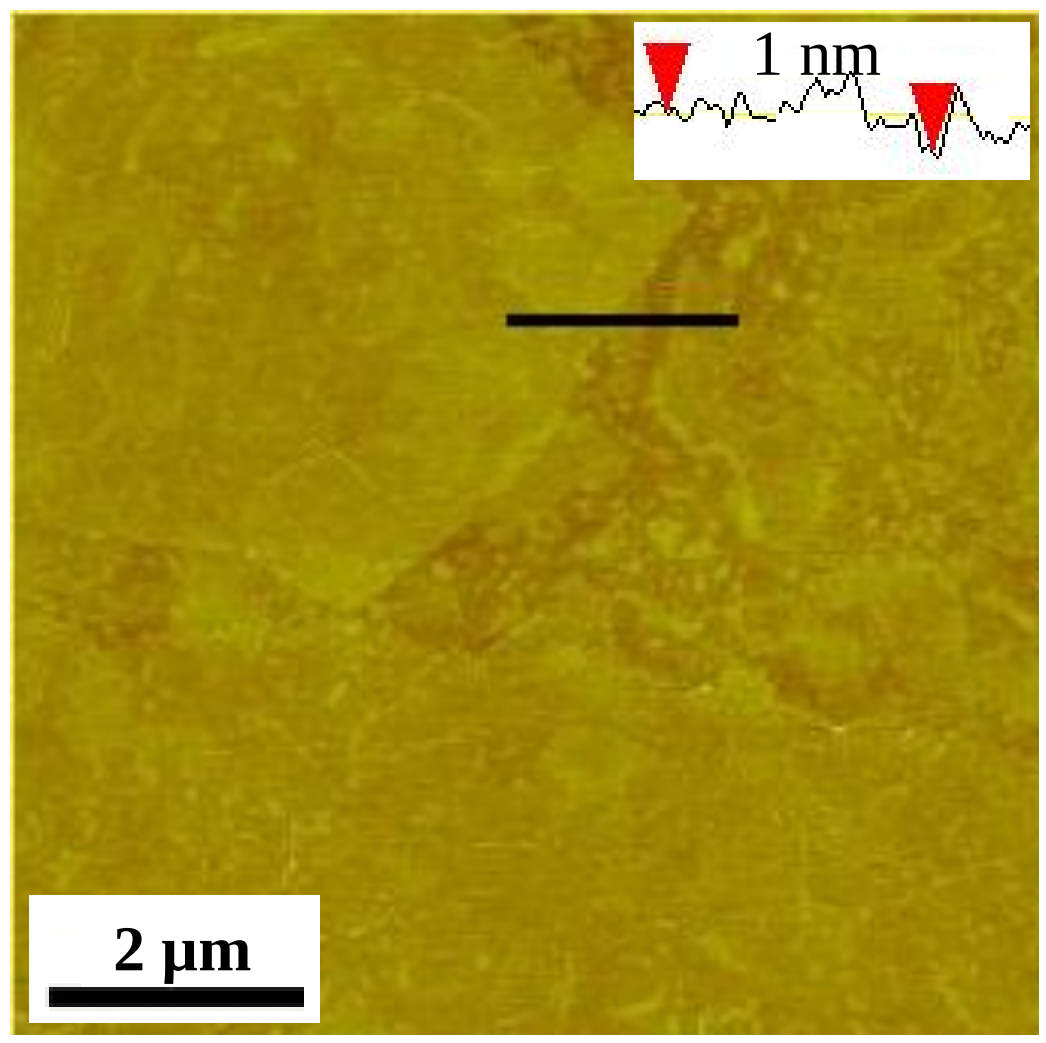

Figure S3. Typical AFM images of 3AA/GO/3AA structure after heat treatment at 200 °C.

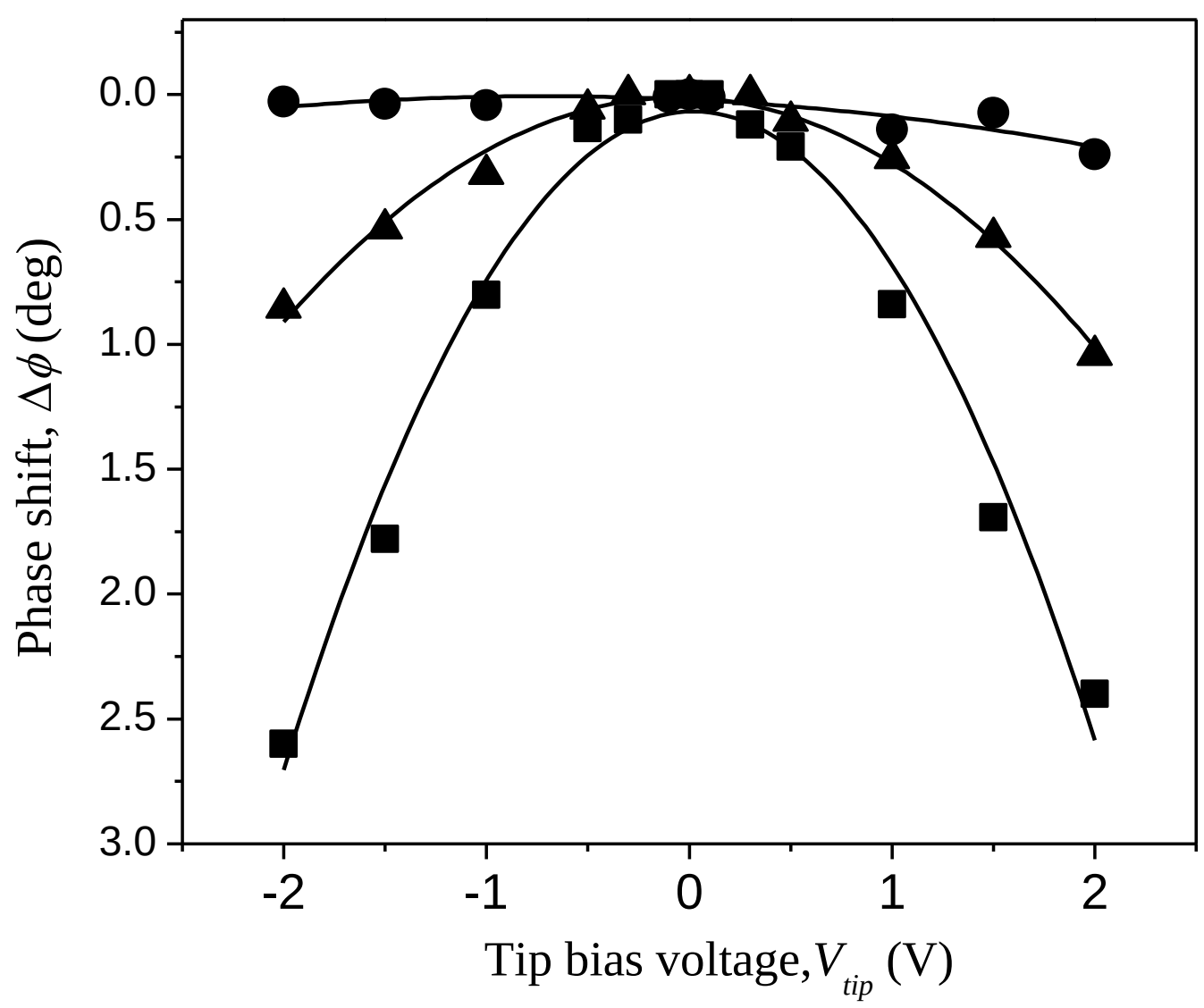

Figure S4. Typical parabolic dependence of $\Delta\Phi$ on tip bias voltage for $h$ = 20 nm at a typical location near the edges of GO (-●-) RGO-1 (-▲-) and RGO-2 (-■-) sheets.

Table S1. Peak positions (in eV) of de-convoluted components of C-1*s* core level XPS spectra of AA/GO/AA and 3AA/GO/3AA structures after heat treatment, as indicated. The integrated intensity values (%) are given in parenthesis. The last column gives the values of O/C ratio and X (ratio of non-graphitic carbon to graphitic carbon).

| Sample (Heat treatment Temperature) | $sp^2$-C | $sp^3$-C | C-O | C=O | COOH | $\pi$-$\pi^*$ | O/C ratio | X |
|---|---|---|---|---|---|---|---|---|
| AA/GO/AA (400 °C) | 284.4 (70) | 285.4 (14) | 286.2 (8) | 287.2 (4) | 288.5 (2) | 289.5 (2) | 0.16 | 0.38 |
| 3AA/GO/3AA (200 °C) | 284.4 (66) | 285.4 (16) | 286.2 (8) | 287.2 (4) | 288.5 (3) | 289.5 (3) | 0.18 | 0.44 |